%% file: DAFx24_tmpl.tex
\def\papertitle{Towards Efficient Modelling of String Dynamics: A Comparison of State Space and Koopman based Deep Learning Methods}
\def\paperauthorA{Rodrigo Diaz}
\def\paperauthorB{Carlos De La Vega Martin}
\def\paperauthorC{Mark Sandler}
\documentclass[twoside,a4paper]{article}
\usepackage{etoolbox}
\usepackage{dafx_24}
\usepackage{amsmath,amssymb,amsfonts,amsthm}
\usepackage{euscript}
\usepackage[T1]{fontenc}
\usepackage[utf8]{inputenc}
\usepackage{ifpdf}
\usepackage[english]{babel}
\usepackage{caption}
\usepackage{color}
\usepackage{soul}
\usepackage{booktabs}
\usepackage{enumitem}
\usepackage{bm}
\usepackage{subcaption}
\usepackage{siunitx}
\usepackage{listings}
\usepackage{tabularx}

\newcommand{\im}[1]{\includegraphics[width=0.235\textwidth]{#1}}

\DeclareMathOperator{\diag}{diag}
\newcommand{\ols}[1]{\mskip.5\thinmuskip\overline{\mskip-.5\thinmuskip {#1} \mskip-.5\thinmuskip}\mskip.5\thinmuskip} % overline short
\newcommand{\olsi}[1]{\,\overline{\!{#1}}} % overline short italic
\makeatletter
\newcommand\closure[1]{
  \tctestifnum{\count@stringtoks{#1}>1} %checks if number of chars in arg > 1 (including '\')
  {\ols{#1}} %if arg is longer than just one char, e.g. \mathbb{Q}, \mathbb{F},...
  {\olsi{#1}} %if arg is just one char, e.g. K, L,...
}
\long\def\count@stringtoks#1{\tc@earg\count@toks{\string#1}}
\long\def\count@toks#1{\the\numexpr-1\count@@toks#1.\tc@endcnt}
\long\def\count@@toks#1#2\tc@endcnt{+1\tc@ifempty{#2}{\relax}{\count@@toks#2\tc@endcnt}}
\def\tc@ifempty#1{\tc@testxifx{\expandafter\relax\detokenize{#1}\relax}}
\long\def\tc@earg#1#2{\expandafter#1\expandafter{#2}}
\long\def\tctestifnum#1{\tctestifcon{\ifnum#1\relax}}
\long\def\tctestifcon#1{#1\expandafter\tc@exfirst\else\expandafter\tc@exsecond\fi}
\long\def\tc@testxifx{\tc@earg\tctestifx}
\long\def\tctestifx#1{\tctestifcon{\ifx#1}}
\long\def\tc@exfirst#1#2{#1}
\long\def\tc@exsecond#1#2{#2}
\makeatother

\input glyphtounicode
\ninept

\newcounter{numauth}
\newcounter{listcnt}
\newcommand\authcnt[1]{\ifdefined#1 \stepcounter{numauth} \fi}

\newcommand\addauth[1]{
\ifdefined#1 
\stepcounter{listcnt}
\ifnum \value{listcnt}<\value{numauth}
\appto\authorslist{, #1}
\else
\appto\authorslist{~and~#1}
\fi
\fi}
\authcnt{\paperauthorB}
\authcnt{\paperauthorC}
\authcnt{\paperauthorD}
\authcnt{\paperauthorE}
\authcnt{\paperauthorF}
\authcnt{\paperauthorG}
\authcnt{\paperauthorH}
\authcnt{\paperauthorI}
\authcnt{\paperauthorJ}
\def\authorslist{\paperauthorA}
\addauth{\paperauthorB}
\addauth{\paperauthorC}
\addauth{\paperauthorD}
\addauth{\paperauthorE}
\addauth{\paperauthorF}
\addauth{\paperauthorG}
\addauth{\paperauthorH}
\addauth{\paperauthorI}
\addauth{\paperauthorJ}
\usepackage{times}
\newif\ifpdf
\ifx\pdfoutput\relax
\else
   \ifcase\pdfoutput
      \pdffalse
   \else
      \pdftrue
\fi

\ifpdf % compiling with pdflatex
  \usepackage[pdftex,
    pdftitle={\papertitle},
    pdfauthor={\authorslist},
    pdfsubject={Proceedings of the 27th International Conference on Digital Audio Effects (DAFx24)},
    colorlinks=false, % links are activated as color boxes instead of color text
    bookmarksnumbered, % use section numbers with bookmarks
    pdfstartview=XYZ % start with zoom=100% instead of full screen; especially useful if working with a big screen :-)
  ]{hyperref}
  \usepackage[pdftex]{graphicx}
\else % compiling with latex
  \usepackage[dvips]{epsfig,graphicx}
  \usepackage[dvips,
    pdftitle={\papertitle},
    pdfauthor={\authorslist},
    pdfsubject={Proceedings of the 27th International Conference on Digital Audio Effects (DAFx24)},
    colorlinks=false, % no color links
    bookmarksnumbered, % use section numbers with bookmarks
    pdfstartview=XYZ % start with zoom=100% instead of full screen
  ]{hyperref}
\fi
\usepackage[hypcap=true]{caption}
\title{\papertitle}

\affiliation{
\paperauthorA \qquad \paperauthorB \qquad \paperauthorC \,
\sthanks{This work was funded by UKRI and EPSRC as part of the ``UKRI CDT in Artificial Intelligence and Music'', under grant EP/S022694/1.}
}
{\href{https://dafx24.surrey.ac.uk}{Centre for Digital Music} \\ Queen Mary University of London, UK 
\\
{
\tt
\href{mailto:r.diazfernandez@qmul.ac.uk}{r.diazfernandez} \quad \href{mailto:c.delavegamartin@qmul.ac.uk}{c.delavegamartin} \quad
\href{mailto:mark.sandler@qmul.ac.uk}{mark.sandler@qmul.ac.uk}}
}

\begin{document}

\graphicspath{{images/}{images/rebuttal/}}

% more pdf-tex settings:
\ifpdf % used graphic file format for pdflatex
  \DeclareGraphicsExtensions{.png,.jpg,.pdf}
\else  % used graphic file format for latex
  \DeclareGraphicsExtensions{.eps}
\fi

%\makeatletter
%\pdfbookmark[0]{\@pdftitle}{title}
%\makeatother

\maketitle
\begin{abstract}
\sloppy
This paper presents an examination of State Space Models (SSM) and Koopman-based deep learning methods for modelling the dynamics of both linear and non-linear stiff strings. Through experiments with datasets generated under different initial conditions and sample rates, we assess the capacity of these models to accurately model the complex behaviours observed in string dynamics. Our findings indicate that our proposed Koopman-based model performs as well as or better than other existing approaches in non-linear cases for long-sequence modelling.

We inform the design of these architectures with the structure of the problems at hand. Although challenges remain in extending model predictions beyond the training horizon (i.e., extrapolation), the focus of our investigation lies in the models' ability to generalise across different initial conditions within the training time interval.
This research contributes insights into the physical modelling of dynamical systems (in particular those addressing musical acoustics) by offering a comparative overview of these and previous methods and introducing innovative strategies for model improvement. Our results highlight the efficacy of these models in simulating non-linear dynamics and emphasise their wide-ranging applicability in accurately modelling dynamical systems over extended sequences.
\fussy
\end{abstract}

\input{01_introduction}

\input{02_background}
\input{03_method}

\input{04_experiments}

\input{05_conclusion}
\input{06_acknowlegments}

% \newpage
% \nocite{*}
\bibliographystyle{IEEEbib}
\bibliography{references} % requires file DAFx24_tmpl.bib

% \cleardoublepage
% \input{rebuttal_section}

\end{document}

%% file: 01_introduction.tex
\section{Introduction}
\label{sec:intro}

The analysis and modelling of acoustic dynamical systems have been critical areas of scientific research and practical applications for many years. Numerous numerical methods, including traditional approaches such as finite differences and finite elements, have been developed to simulate a broad spectrum of acoustic phenomena within dynamic systems. These methods provide robust frameworks for simulation but often encounter complexities and limitations, particularly with nonlinear systems or those requiring real-time interaction~\cite{bilbao_numerical_2009}.

Learning-based methods are becoming increasingly prevalent due to their potential to model complex dynamics efficiently. Such methods have been proposed to address some of these obstacles in physical modelling. In particular, neural operator-based architectures~\cite{li_fourier_2020} and and their recurrent variants~\cite{parker_physical_2022} have been explored for this purpose. However, in such cases, training sequences are limited to dozens or hundreds of steps, with performance degrading rapidly when extrapolating beyond the training domain~\cite{delavegamartin_physical_2023, michalowska_neural_2023}. These methods face difficulties in training longer sequences, primarily due to gradient instability resulting from vanishing or exploding gradients~\cite{goodfellowDeepLearning2016}. Additionally, using nonlinearities in the recursion considerably slows down the training process~\cite{orvieto_resurrecting_2023}.

In contrast, optimising linear layers is computationally more efficient, as the recursion can be parallelised more effectively~\cite{smith_simplified_2022}. The emergence of deep learning methods that incorporate linear dynamics offers an alternative perspective for modelling problems of this kind. In particular, methods grounded in Koopman theory have shown potential in accurately capturing the nuances of non-linear dynamic systems~\cite{lusch_deep_2018}. Similarly, deep SSMs have achieved success in efficiently processing very long sequences across various tasks~\cite{gu_efficiently_2022}.

This preliminary work applies these methods to simulate the dynamics of one-dimensional strings and provides a comprehensive comparison of several related neural architectures to model such dynamics. We introduce a modified Koopman-based model that can capture the nonlinear dynamics of strings with greater accuracy and fewer parameters than previous methods~\cite{parker_physical_2022}. We also investigate the connections between analytical solutions to these challenges and the architectural decisions behind these deep learning models, providing insights into their effectiveness.

% While it is possible to use a more traditional technique to model such dynamics efficiently~\cite{avanzini_efficient_2012}, we use neural methods  its potential for more complex use cases.

%% file: 02_background.tex
% \vspace{-4pt}
\section{Background}
\label{sec:background}
% \vspace{-4pt}
\subsection{The stiff string}
The motion of a stiff string defined over the interval $x \in [0, \ell]$ and with no external forcing can be expressed as
\begin{equation}
\rho A \frac{\partial^2 y}{\partial t^2} + EI \frac{\partial^4 y}{\partial x^4} - T(y) \frac{\partial^2 y}{\partial x^2} + d_1 \frac{\partial y}{\partial t} - d_3 \frac{\partial^3 y}{\partial x^2 \partial t} = 0,
\end{equation}
Where $y(x,t)$ is the transverse displacement of the string, $\rho$ is the mass density, $A$ is the cross-sectional area, $E$ is the Young's modulus, $I$ is the second moment of area, $T(y)$ is the displacement dependent tension, and $d_1$ and $d_3$ are the damping coefficients. 
This PDE, together with a valid set of boundary conditions and initial conditions, defines the Initial-Boundary Value Problem (IBVP) that we are interested in solving. (see Table~\ref{tab:parameters} for the parameter values used).

\subsubsection{Linear case}
\label{sssec:linear}
For small displacements, we can take $T(y) = T_0$ to be a constant. Assuming simply-supported boundary conditions, $y(0,t)=y(\ell,t)=0$ and $y^{\prime \prime}(0,t)=y^{\prime \prime}(\ell,t)=0$, and using separation of variables, we can arrive to a set of spatial basis functions $K_\mu(x) = \sin \left(\frac{\mu \pi x}{\ell}\right)$ for $\mu =1, 2, \dots$. These basis functions, which are real in this specific case, diagonalise the system~\cite{trautmann_sound_2000, avanzini_efficient_2012}: 
\begin{equation}
    y(x,t) = \sum_{\mu=1}^{\infty} \frac{\bar{y}_\mu(t) K_\mu(x)}{\left\|K_\mu(x)\right\|_2^2}.
\label{eq:basis}
\end{equation}

The spatial modal shapes $K_\mu(x)$ constitute the columns of our transformation kernel $\bm{K}(x, \mu)$ of the associated Sturm-Liouville transform (SLT).

\subsubsection{Tension-modulated nonlinearity}
\label{sssec:nonlinear}

A common assumption when incorporating nonlinear behaviour is to assume spatially uniform tension modulation, leading to the Kirchhoff-Carrier equation~\cite{carrierNonlinearVibrationProblem1945} with the tension expressed as:
\begin{equation}
    \begin{aligned}
        T_{N L}(y(x, t)) & =T_0+T_1(y(x, t)) \\
        & =T_0+\frac{E A}{2 \ell} \int_0^{\ell} y^{\prime 2}(x, t) d x.
    \end{aligned}
\end{equation}
Our transformation kernel $\bm{K}(x, \mu)$ remains the same as in the linear case because our assumption of homogeneous tension still holds~\cite{trautmann_sound_2000}.
Applying the transformation kernel to the non-linear PDE, we obtain a set of ODEs in modal space:
\begin{equation}
    \begin{aligned}
        & \rho A \ddot{\bar{y}}(\mu, t)+ \left(d_3 \eta_\mu^2+d_1\right) \dot{\bar{y}}(\mu, t)+ \\
        & + (E I \eta_\mu^4 + T_0 \eta_\mu^2) \bar{y}(\mu, t)-\bar{b}(\mu, y, \bar{y})=0,
    \end{aligned}
    \label{eq:nonlinearodes}
\end{equation}
where the non-linear term $\bar{b}(\mu, y, \bar{y})$ in modal space  is given by:
\begin{equation}
    \bar{b}(\mu, y, \bar{y})=\eta_\mu^2\left(\frac{E A}{2 \ell} \int_0^{\ell} (y^{\prime})^2(\xi) d \xi\right) \bar{y}(\mu).    
\end{equation}
These equations can be solved by discretising in time and space and using numerical methods such as the Runge-Kutta method~\cite{cheneyNumericalMathematicsComputing2012}.

\subsection{Discrete-time dynamical systems}

The dynamics of the string can be modelled using the general description of discrete-time dynamical systems
\begin{equation}
\bm{x}_{k+1} = \bm{F}(\bm{x}_k),
\end{equation}
where $\bm{x} \in \mathbb{R}^n$ are state variables at discrete time $k$, and $\bm{F}$ is a non-linear operator that advances the state.
In the linear case, the evolution of the state is given by the state transition matrix $\bm{A}$, where $\bm{\dot{x}} = \bm{A}\bm{x}$ and therefore, $\bm{x}_{k+1} = e^{\bm{A} \Delta t} \bm{x}_k$ with a chosen sampling period $\Delta t$. The linear operator $\bm{A}$ for the system can be diagonalised using the similarity transformation
\begin{equation}
    \bm{A} = \bm{E^{-1}} \bm{\tilde{A}} \bm{E},
    \label{eq:similarity}
\end{equation}
where $\bm{E}\in \mathbb{C}^{m \times n}$ and $\bm{E}^{-1}$ correspond to the inverse and forward SLT~\cite{schafer_simulation_2020} kernel. $\bm{\Tilde{A}} \in \mathbb{C}^{m \times m}$ is a diagonal matrix of eigenvalues.
The state variables are $\bm{x} = \begin{bmatrix} \bm{u} & \bm{v} \end{bmatrix}^T$, where $\bm{u}$ and $\bm{v}$ are the deflection and velocity, respectively.
$\bm{E}$ corresponds to $\bm{K}(x, \mu)$ of Section \ref{sssec:linear} in the case of the linear stiff string.

In the non-linear case, the state evolution is given by the non-linear operator $\bm{F}$ and, in the general case, it cannot be diagonalised. This motivates the use of deep learning methods. For example, a neural network can be designed to learn the dynamics of the non-linear system by adopting a structure that mirrors this problem formulation.

We consider two methods to learn the dynamics of this system. The first method draws inspiration from Koopman theory and consists of three components: one that transforms the input coordinates (such as position and velocity), a middle component that advances the linearised dynamics through time, and a final component that transforms the states back to the original coordinate system.

The second method relies on deep SSMs. Here, the same three functional components identified in the Koopman-inspired method are integrated into a single layer. By stacking multiple such layers, incorporating nonlinear activations, and adding skip connections, we facilitate the learning of the system's nonlinear dynamics.

\subsection{The Koopman operator and the Dynamic Mode Decomposition}

Learning the Koopman operator aims to derive a finite approximation to an infinite-dimensional linear operator. This approach essentially transforms the problem of modelling non-linear dynamics into one of linear dynamics in the space of observables~\cite{williams_data_2015}. Similarly, Dynamic Mode Decomposition (DMD) can identify and approximate non-linear dynamics using a linear operator~\cite{tu_dynamic_2014, brunton_modern_2022}. DMD is used to find the best fit, in the least squares sense, for the matrices in Eq.~\ref{eq:similarity} efficiently, even for cases where the state variable vector $\bm{x}_k$ is very tall. The dynamics under the Koopman operator $\mathcal{K}g = g \circ F$, where the $\circ$ denotes function composition, evolve as:

\begin{equation}
g\left(\bm{x}_{k+1}\right)=\mathcal{K} g\left(\bm{x}_k\right) .
\end{equation}

The dynamics found with DMD are equivalent to the Koopman operator when the measurement functions $g(\bm{x})$ are linear~\cite{brunton_modern_2022}. That is, for systems such as the linear stiff string, it is possible to extract the eigenfunctions $\bm{K}$ and the eigenvalues in $\bm{\Tilde{A}}$ directly from discrete observations of the system. 

\subsection{Linear Recurrent Unit and State Space Models}

State-space models (SSMs) describe a continuous-time dynamical system 
\begin{equation}
\bm{\dot{x}}(t) = \bm{A x}(t)+\bm{B u}(t), \quad \bm{y}(t)=\bm{C x}(t)+\bm{D u}(t),    
\end{equation}
where $\bm{x}(t)$ is the internal state as a function of time, $\bm{A}$ again is the transition matrix that advances the dynamics of the system. $\bm{C}$, $\bm{B}$ transform the input coordinates into state space and viceversa, $\bm{D}$ is a feed-through matrix (similar to a skip connection) and $\bm{y}(t)$ is the output. In practice, the SSM is often diagonalised~\cite{gupta_diagonal_2022}, using a similarity transformation (Eq.~\ref{eq:similarity}) so that we obtain $\bm{\Tilde{A}}$, in addition to $\bm{\Tilde{B}} = \bm{E}^{-1}\bm{B}$ and $\bm{\Tilde{C}} = \bm{C}\bm{E}$.
% \begin{align}
% \bm{x}_{k+1} &= \bm{\tilde{A}}\bm{x}_k + \bm{\tilde{B}}\bm{u}_k \\
% \bm{y}_k &= \bm{\tilde{C}}\bm{x}_k + \bm{D}\bm{u}_k
% \end{align}

During training, the matrices are discretised using bilinear or Zero-Order Hold (ZOH) methods with step parameter $\Delta$. Although these matrices can generally vary over time, deep SSMs~\cite{gu_efficiently_2022} and its diagonal variants (DSSMs)~\cite{gupta_diagonal_2022}) aim to learn the parameters for linear time-invariant systems. More recently, there have been newer techniques to facilitate the optimisation of time-varying parameters~\cite{smith_simplified_2022}. The latter methods (S5) as well as the structured DSSMs initialise these matrices with a predefined structure derived from HiPPO theory~\cite{gu_how_2022}.

The model then optimises $\diag(\bm{\Tilde{A}}) \in \mathbb{C}^{n}$ along with the $\bm{\tilde{B}} \in \mathbb{C}^{m \times n}$, $\bm{\tilde{C}} \in \mathbb{C}^{n \times m}$ and $\diag({\bm{D}}) \in \mathbb{R}^n$ matrices (the matrix feed-through matrix ${\bm{D}}$ is set to be diagonal). Furthermore, the discretisation step $\Delta$ is also learnt. The time-invariant DSSM can be optimised efficiently~\cite{gu_parameterization_2022} through the linear convolution of
\begin{align}
\bm{y} &= \closure{\bm{K}} \ast \bm{u} \\
\closure{\bm{K}} &= \left(\closure{\bm{B}}^{\top} \odot \bm{C}\right) \cdot \mathcal{V}_L(\closure{\bm{A}}),
\end{align}
where and the matrices $\closure{\bm{B}}$ and $\closure{\bm{A}}$ are the discretised versions of $\bm{\tilde{B}}$ and $\bm{\tilde{A}}$, $\mathcal{V}_L$ is the Vandermonde matrix of size $L$, and $\odot$ is the element-wise product. Alternatively, in S5 the diagonal matrices are also efficiently optimised using a parallel scan~\cite{blelloch_prefix_1990} to parallelise the recursion.

The linear recurrent unit (LRU)~\cite{orvieto_resurrecting_2023} follows a very similar formulation but differs in its initialisation strategy. Instead of structured initialisation, the vector $\diag(\bm{\tilde{A}})$ is initialised with eigenvalues drawn from a uniform distribution within specified bounds in the unit circle.

%% file: 03_method.tex
% \vspace{-4pt}
\section{Method}
\label{sec:method}
% \vspace{-4pt}

To learn the linear and non-linear string dynamics we focus on a Koopman based model with time-varying state.

\subsection{Koopman Model}

The Koopman-based model leverages an autoencoder framework, in which both the encoder and decoder are designed as multilayer perceptrons (MLPs), to learn the system dynamics in the continuous domain, inspired by previous work~\cite{lusch_deep_2018, brunton_modern_2022}. The encoder, denoted by \( \varphi \), transforms the state of the system into an observable embedding, which is then evolved in time according to the linear operator \( \mathcal{K} \):

\begin{equation}
\varphi\left(\mathbf{x}_{k+1}\right)=\mathcal{K} \varphi\left(\mathbf{x}_k\right).
\end{equation}

Here, the encoder and decoder approximate the Koopman eigenfunctions, which, in the context of a linear stiff string, act as approximate eigenfunctions of the SLT transform.

The main idea for the middle component is to optimise the eigenvalues contained within \( \mathbf{\Lambda} = \diag(\mathbf{\tilde{A}}) \), ensuring that:
\begin{equation}
\mathbf{x}_{k+1} = \varphi^{-1}(\mathbf{\Lambda}_{\nu, \theta} \varphi(\mathbf{x}_k)),
\end{equation}
where the eigenvalues \( \mathbf{\Lambda} = \begin{bmatrix} \lambda_1, \ldots, \lambda_n \end{bmatrix}^\mathsf{T} \) are parameterised by their real and imaginary components, \( \nu = \Re(\lambda) \) and \( \theta = \Im(\lambda) \), in the continuous domain.

Optimisation of the eigenvalues can be approached through recursion or frequency sampling. Learning the linear dynamics of the system in frequency equates to optimising the resonant poles of a parallel filter bank, as per the frequency sampling method~\cite{colonel_direct_2022, diaz_rigidbody_2023}. Each filter within this bank acts as a two-pole resonator.

Given the linear dynamics of the model, naive recursion can be markedly slow. Using the parallel scan approach, as employed in the S5 model, can significantly enhance the efficiency of recursion. This technique also facilitates modelling of time-varying dynamics that cannot be captured by static methods such as frequency sampling or convolution. The latter is a helpful feature to capture the time-varying spectra present in systems such as the tension-modulated string~\cite{lusch_deep_2018}. Following this reasoning, we add an MLP that acts on the state of the system, where the input is the square of the real and imaginary components of the state.

For improved optimisation, the parameters are optimised independently, using exponential parameterisation to ensure stability~\cite{orvieto_resurrecting_2023}. The magnitude and phase are optimised in logarithmic space, thereby updating the eigenvalue \( \lambda_d \) with
\begin{equation}
\lambda_d := \exp(-\exp(\nu) + i\theta),
\end{equation}
indicating that optimisation involves $\log(\nu)$ and $\log(\theta)$. This approach mirrors the parameterisation strategies utilised in SSMs and aligns with the design principles adopted for the LRU.

\subsection{SSMs with initial conditions}
\label{sec:ssm_ic}

While the formulation of state space models is suited for mapping $\bm{u} \mapsto \bm{y}$ (many-to-many), it can adapted to map an initial condition to a sequence $\bm{u}_0 \mapsto \bm{u_{1:L}}$ (one-to-many). Thus, the LRU and the S5 model have been adapted to learn a first layer that produces a sequence from initial conditions.
\begin{equation}
\bm{u_{k+1}} = \closure{\bm{A}} \bm{x}_k + \closure{\bm{B}}  \bm{u}_k \delta_k, \quad \bm{y}_k =\closure{\bm{C}} \bm{x}_k,  
\end{equation}
where $\bm{u}$ has the initial conditions of the system at $k=0$ and $\delta$ is a unit impulse function. In this setup, the output $\bm{y}$ must be delayed by one sample to correspond to the samples after the initial condition. Because of this, once trained, it is straightforward to use an arbitrary excitation signal with the SSM model at inference time. Alternatively, the states for the rest of the sequence can be obtained by multiplying the initial condition with the Vandermonde matrix 
$\bm{x}_{1:L} = \nu_L(\closure{\bm{A}}) \bm{x}_0$.

For the linear stiff string, it is necessary, in principle, to optimise only a single layer of such a model. However, the non-linear approximation requires stacking many such layers with non-linear activations and skip connections.

%% file: 04_experiments.tex
% \vspace{-4pt}
\section{Experiments}
\label{sec:experiments}
% \vspace{-4pt}

\begin{figure*}[ht]
  \centering
  \begin{subfigure}[b]{0.9\textwidth}
    \includegraphics[width=\linewidth]{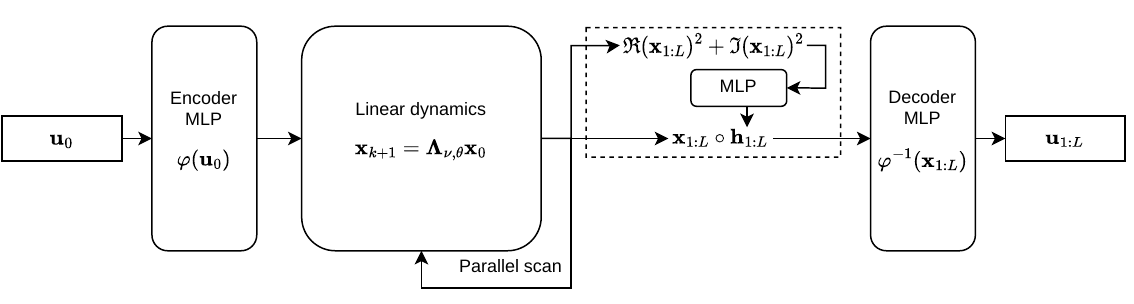}
    \caption{}
    \label{fig:a}
  \end{subfigure}
  \caption{Koopman-based architecture, with optional varying processing for the generated states, as indicated within the dashed box. This involves an MLP, where the eigenvalue radii serve as input, and its output is applied to the state sequence through element-wise multiplication.}
\label{fig:models}
\end{figure*}

\subsection{Dataset}
\label{ssec:dataset}

\begin{table}[b]
\centering
\begin{tabular}{
  l
  S[table-format=1.4e-2] % Adjust format for numbers
  l
}
\toprule
{\textbf{Parameter}} & {\textbf{Value}} & {\textbf{Unit}} \\
\midrule
$A$ & 0.5188e-6 & \si{\meter\squared} \\
$I$ & 0.141e-12 & \si{\meter\tothe{4}} \\
$\rho$ & 1140 & \si{\kilogram\per\meter\cubed} \\
$E$ & 5.4e9 & \si{\pascal} \\
$d_1$ & 8.0e-5 & \si{\kilogram\per\meter\per\second} \\
$d_3$ & 1.4e-5 & \si{\kilogram\meter\per\second} \\
$T$ & 60.97 & \si{\newton} \\
$\ell$ & 0.65 & \si{\meter} \\
\bottomrule
\end{tabular}
\caption{Physical parameters of a nylon guitar string, tuned to $\bm{B}_3$.}
\label{tab:parameters}
\end{table}

Our experiments were carried out on two types of string models: the linear stiff string and the tension modulated stiff string. We generated a dataset of $L = 4000$ time steps for each model, using sample rates of $4\si{\kHz}$ and $16\si{\kHz}$. For each sample rate we created 2 sets of 1,000 distinct initial conditions, with either Gaussian or noise-like initial displacement and zero initial velocity. The noise-like initial conditions were randomly generated based on a uniform distribution with the maximum displacements ranging from $1\si{\mm}$ to $1\si{\cm}$. The Gaussian-like initial conditions were determined by randomly selected means and standard deviations, also with the same maximum initial displacement range. The dataset was generated by solving the tension-modulated string Eq.~\ref{eq:nonlinearodes} numerically using an explicit Runge-Kutta solver of order 8(5,3) (\lstinline{DOP853})~\cite{hairer_rungekutta_1993}. Although both position and velocity were generated for each time step, only the displacement data was used as input for the models. The simulation data was normalised to ensure a standard deviation of 1 prior to training. 80\% of the dataset was used for training and another 10\% was used for validation.

The selection of these parameters accurately emulates the characteristics of a nylon guitar string tuned to $\bm{B}_3$ and was in alignment with the experimental setups referenced in previous studies~\cite{schafer_string_2020,rabenstein_digital_2003}. This approach ensures that our investigations are grounded in realistic parameters, as defined in Table~\ref{tab:parameters}.

\subsection{Models}

We compare six models: an autoencoder without varying state (Koopman), an autoencoder with with varying state (Koopman\textsubscript{var}), an LRU model~\cite{orvieto_resurrecting_2023}, an S5 model~\cite{smith_simplified_2022}, and the Fourier Neural Operator embedded in the RNN (FRNN) and GRU (FGRU) architectures~\cite{parker_physical_2022}. 

The Koopman models and SSMs were trained for 2500 epochs with early stopping, using the AdamW optimiser~\cite{loshchilov_decoupled_2018} with a constant learning rate of 0.003. The FNO models were trained for 2000 epochs, using the same optimiser with a 1-cycle learning-rate scheduling ranging from $10^{-4}$ to $10^{-3}$. The hyperparameters for the S5 model were adapted from those used in a speech classification task, with the adjustment of disabling bidirectionality for our specific use case. Similarly, the LRU model employed comparable parameters, but with 8 layers instead of the 6 used for S5. Additionally, the initial bounds for the eigenvalues (for the Koopman and LRU models) are specifically set between 0.99 and 1 within the unit circle. The parameters for the FNO models were the same as those used in previous similar experiments~\cite{parker_physical_2022}, except for the batch size, which was reduced to 256 due to memory limitations caused by the longer sequence (400 time steps instead of the original 120 used for training). 

The Koopman model is structured with three dense layers each for the encoder and decoder. The Koopman\textsubscript{var} employs a single dense layer for input and output, simplifying the network architecture. Both the SSMs and Koopman-based models internal state representation has size 128, ensuring consistent dimensionality across models.

The models are implemented using JAX. We base our models on the original implementation of S5~\footnote{\url{https://github.com/lindermanlab/S5}} and an unofficial implementation of LRU~\footnote{\url{https://github.com/NicolasZucchet/minimal-LRU}}. The FGRU and FRNN models are based on the official PyTorch implementation~\footnote{\url{https://github.com/julian-parker/DAFX22_FNO}}. An overview of the Koopman model architecture is provided in Fig.~\ref{fig:models} and the number of parameters for each model can be found in Table~\ref{tab:trainable_parameters}. 

The training loss for the SSM models is based on the mean square error (MSE) and for the FNO models is the $log_{10}$ MSE, whereas the Koopman models use a composite loss comprising MSE for signal prediction, encoding loss, and consistency loss (also called linear dynamics loss~\cite{lusch_deep_2018}):

\begin{align}
\mathcal{L_{\textit{consistency}}} &= \sum_{k=1}^{L - 1} \|\varphi\left(\bm{x}_{k}\right)-\Lambda^k \varphi\left(\bm{x}_0\right)\|_2^2 \\[0.5ex]
\mathcal{L_{\textit{pred}}} &= \sum_{k=1}^{L - 1} \| \bm{x}_k - \varphi^{-1}(\Lambda^k \varphi (\bm{x}_0)) \|_2^2 \\[0.5ex]
\mathcal{L_{\textit{enc}}} &= \| \bm{x}_0 - \varphi^{-1}(\varphi(\bm{x}_0)) \|_2^2 \\[1ex]
\mathcal{L} &= \alpha_1 \mathcal{L_{\textit{pred}}} + \alpha_2 \mathcal{L_{\textit{enc}}} + \alpha_3 \mathcal{L_{\textit{consistency}}}.
\end{align}

The encoding loss ensures that the decoder accurately reverses the encoder's transformation, while the consistency loss enforces that the observed state evolution and the projection operation are commutative. The loss weights were set to $\alpha_1 = 1.0$, $\alpha_2 = 1.0$, and $\alpha_3 = 0.01$.

\begin{table}[htbp]
\centering
\begin{tabular}{lc}
\toprule
Model & Trainable Parameters \\
\midrule
Koopman & 191,077 \\
Koopman\textsubscript{var} & 134,885 \\
LRU & 636,221 \\
S5 & 307,879 \\
FGRU & 640,033 \\
FRNN & 643,201 \\
\bottomrule
\end{tabular}
\caption{Number of trainable parameters for each model.}
\label{tab:trainable_parameters}
\end{table}

\begin{table*}[!ht]
\begin{subtable}{1\textwidth}
    \centering
    \vspace{-4pt}
    \resizebox{\linewidth}{!}{
    \begin{tabular}{l c c c c c c c c}
        \toprule
         & \multicolumn{4}{c}{\textbf{Gaussian}} & \multicolumn{4}{c}{\textbf{Noise}} \\
         \cmidrule(lr){2-5} \cmidrule(lr){6-9}
         & \multicolumn{2}{c}{\textbf{4000kHz}} & \multicolumn{2}{c}{\textbf{16000kHz}} & \multicolumn{2}{c}{\textbf{4000kHz}} & \multicolumn{2}{c}{\textbf{16000kHz}} \\
         \cmidrule(lr){2-3} \cmidrule(lr){4-5} \cmidrule(lr){6-7} \cmidrule(lr){8-9}
        \textbf{Model}            & \textbf{MSE Rel} & \textbf{MAE Rel} & \textbf{MSE Rel} & \textbf{MAE Rel} & \textbf{MSE Rel} & \textbf{MAE Rel} & \textbf{MSE Rel} & \textbf{MAE Rel} \\ 
        \midrule
        DMD                       & 1.1865(0.2019)           & 0.8289(0.0857)          & 1.0938(0.1311)          & 0.8747(0.0749)          & 1.6695(0.2687)          & 1.0420(0.1178)          & 1.8313(0.2467)          & 1.0953(0.087) \\
        Koopman                   & 0.1041(0.0144)           & 0.3271(0.0259)          & 0.0971(0.0029)          & 0.3117(0.0044)          & 0.2070(0.0939)          & 0.3978(0.1084)          & 0.0865(0.0155)          & 0.2596(0.0275) \\
        Koopman\textsubscript{var} & \textbf{0.0113}(0.0061) & \textbf{0.0977}(0.0284) & \textbf{0.0094}(0.0046) & \textbf{0.0888}(0.0245) & \textbf{0.0041}(0.0008) & \textbf{0.0527}(0.0081) & \textbf{0.0531}(0.0572) & \textbf{0.1714}(0.1056) \\
        LRU                       & 0.0260(0.0067)           & 0.1571(0.0240)          & 0.0390(0.0187)          & 0.1870(0.0450)          & 0.0250(0.0065)          & 0.1284(0.0077)          & 0.0476(0.0058)          & 0.1922(0.0125) \\
        S5                        & 0.1165(0.0260)           & 0.3318(0.0291)          & 0.0224(0.0013)          & 0.1467(0.0053)          & 0.0507(0.0127)          & 0.1916(0.0190)          & 0.0387(0.0062)          & 0.1493(0.0078) \\
        \midrule
        \multicolumn{9}{c}{\textit{Models below are trained and evaluated with only 400 time steps}} \\
        \midrule
        DMD & 0.4519(0.0000) & 0.5570(0.0005) & 0.4263(0.0000) & 0.5415(0.0007) & 0.2304(0.0000) & 0.3300(0.0009) & 0.4192(0.0000) & 0.4959(0.0012) \\
        Koopman & 0.0031(0.0005) & 0.0499(0.0050) & 0.0071(0.0002) & 0.0946(0.0022) & 0.0036(0.0007) & 0.0353(0.0052) & 0.0393(0.0013) & 0.1767(0.0020) \\
        Koopman\textsubscript{var} & 0.0008(0.0003) & \textbf{0.0245(0.0054)} & \textbf{0.0003(0.0000)} & \textbf{0.0201(0.0006)} & \textbf{0.0005(0.0000)} & \textbf{0.0132(0.0009)} & \textbf{0.0118(0.0009)} & \textbf{0.0720(0.0022)} \\
        LRU & \textbf{0.0008(0.0000)} & 0.0252(0.0011) & 0.0013(0.0001) & 0.0376(0.0022) & 0.0008(0.0001) & 0.0189(0.0018) & 0.0259(0.0009) & 0.1357(0.0031) \\
        S5 & 0.0124(0.0032) & 0.0910(0.0131) & 0.0131(0.0033) & 0.1145(0.0172) & 0.0057(0.0011) & 0.0672(0.0090) & 0.0357(0.0048) & 0.1721(0.0084) \\
        FGRU & 0.4534(0.0219) & 0.6897(0.0204) & 0.0388(0.0151) & 0.2164(0.0464) & 0.0870(0.0249) & 0.2925(0.0444) & 0.0583(0.0053) & 0.2393(0.0114) \\ 
        FRNN & 0.6927(0.1655) & 0.8489(0.0925) & 0.7636(0.0883) & 0.9311(0.0338) & 0.7289(0.0653) & 0.8553(0.0495) & 0.5611(0.0893) & 0.7623(0.0448) \\         
        \bottomrule \\
    \end{tabular}}
    \vspace{-8pt}
    \caption{Results for the non-linear dataset.}
    \vspace{-0.0cm}
    \label{tab:results}
\end{subtable}

\bigskip
\vspace{-4pt}
\begin{subtable}{1\textwidth}
    \centering
    \vspace{-0.0cm}
    \resizebox{\linewidth}{!}{
    \begin{tabular}{l c c c c c c c c}
        \toprule
         & \multicolumn{4}{c}{\textbf{Gaussian}} & \multicolumn{4}{c}{\textbf{Noise}} \\
         \cmidrule(lr){2-5} \cmidrule(lr){6-9}
         & \multicolumn{2}{c}{\textbf{4000kHz}} & \multicolumn{2}{c}{\textbf{16000kHz}} & \multicolumn{2}{c}{\textbf{4000kHz}} & \multicolumn{2}{c}{\textbf{16000kHz}} \\
         \cmidrule(lr){2-3} \cmidrule(lr){4-5} \cmidrule(lr){6-7} \cmidrule(lr){8-9}
        \textbf{Model}             & \textbf{MSE Rel} & \textbf{MAE Rel} & \textbf{MSE Rel} & \textbf{MAE Rel} & \textbf{MSE Rel} & \textbf{MAE Rel} & \textbf{MSE Rel} & \textbf{MAE Rel} \\ 
        \midrule
        DMD                        & \textbf{0.0000}(0.0000) & \textbf{0.0060}(0.0007) & \textbf{0.0001}(0.0000) & \textbf{0.0097}(0.0006) & \textbf{0.0000}(0.0000) & \textbf{0.0042}(0.0004) & \textbf{0.0001}(0.0000) & \textbf{0.0066}(0.0004) \\
        Koopman                    & 0.0110(0.0069)          & 0.1005(0.0331)          & 0.0048(0.0009)          & 0.0692(0.0074)          & 0.0073(0.0052)          & 0.0789(0.0322)          & 0.0019(0.0001)          & 0.0423(0.0004)  \\
        Koopman\textsubscript{var} & 0.0020(0.0013)          & 0.0427(0.0145)          & 0.0013(0.0012)          & 0.0321(0.0142)          & 0.0002(0.0001)          & 0.0134(0.0023)          & 0.0495(0.0973)          & 0.1125(0.1785)  \\
        LRU                        & 0.0014(0.0005)          & 0.0373(0.0065)          & 0.0010(0.0002)          & 0.0317(0.0034)          & 0.0005(0.0001)          & 0.0199(0.0025)          & 0.0005(0.0002)          & 0.0217(0.0036)  \\
        S5                         & 0.0386(0.0273)          & 0.1896(0.0597)          & 0.0156(0.0102)          & 0.1206(0.0421)          & 0.0112(0.0029)          & 0.1023(0.0139)          & 0.0037(0.0012)          & 0.0539(0.0119)  \\
        \midrule
        \multicolumn{9}{c}{\textit{Models below are trained and evaluated with only 400 time steps}} \\
        \midrule
        DMD & \textbf{0.0000(0.0000)} & \textbf{0.0013(0.0000)} & \textbf{0.0000(0.0000)} & \textbf{0.0006(0.0000)} & \textbf{0.0000(0.0000)} & \textbf{0.0012(0.0000)} & \textbf{0.0000(0.0000)} & \textbf{0.0005(0.0000)} \\
        Koopman & 0.0002(0.0000) & 0.0140(0.0014) & 0.0004(0.0002) & 0.0243(0.0045) & 0.0001(0.0000) & 0.0088(0.0005) & 0.0003(0.0001) & 0.0169(0.0021) \\
        Koopman\textsubscript{var} & 0.0000(0.0000) & 0.0058(0.0019) & 0.0000(0.0000) & 0.0074(0.0018) & 0.0000(0.0000) & 0.0056(0.0006) & 0.0000(0.0000) & 0.0055(0.0004) \\
        LRU & 0.0002(0.0000) & 0.0140(0.0012) & 0.0002(0.0000) & 0.0171(0.0003) & 0.0001(0.0000) & 0.0091(0.0005) & 0.0002(0.0000) & 0.0148(0.0012) \\
        S5 & 0.0017(0.0005) & 0.0425(0.0068) & 0.0043(0.0015) & 0.0616(0.0130) & 0.0017(0.0007) & 0.0363(0.0088) & 0.0108(0.0009) & 0.0944(0.0042) \\
        FGRU & 0.4399(0.0258) & 0.6825(0.023) & 0.0378(0.0162) & 0.214(0.0551) & 0.0681(0.0198) & 0.2618(0.0385) & 0.0470(0.0034) & 0.2145(0.0082) \\ 
        FRNN & 0.6474(0.1711) & 0.8183(0.1198) & 0.7242(0.1912) & 0.8925(0.1281) & 0.5821(0.3888) & 0.6870(0.3413) & 0.4730(0.3205)	 & 0.6347(0.2836) \\         
        \bottomrule \\
    \end{tabular}}
    \vspace{-8pt}
    \caption{Results for the linear dataset.}
    \vspace{-0.0cm}
    \label{tab:results_linear}

\end{subtable}

\caption{Mean and standard deviation (in parentheses) for 5 seeds of a) non-linear and b) linear validation data across different models and sampling rates, under Gaussian and noise-like initial conditions. We use 4000 steps for both $4\si{\kHz}$ and $16\si{\kHz}$. Additionally, the models are also trained and evaluated with 400 time steps to compare with the FRNN and FGRU methods. For DMD, the standard deviation of MSE and MAE across the validation dataset is included as it does not depend on a seed.}
% \vspace{-0.0cm}
\end{table*}

\vspace{-4pt}
\subsection{Results}

\begin{figure*}[!htb]
    \centering
    \setlength{\tabcolsep}{2pt} % Set table column separation
    \renewcommand{\arraystretch}{1} % Set table row separation
    \begin{tabular}{ccc}
        \begin{tabular}{c}
            \im{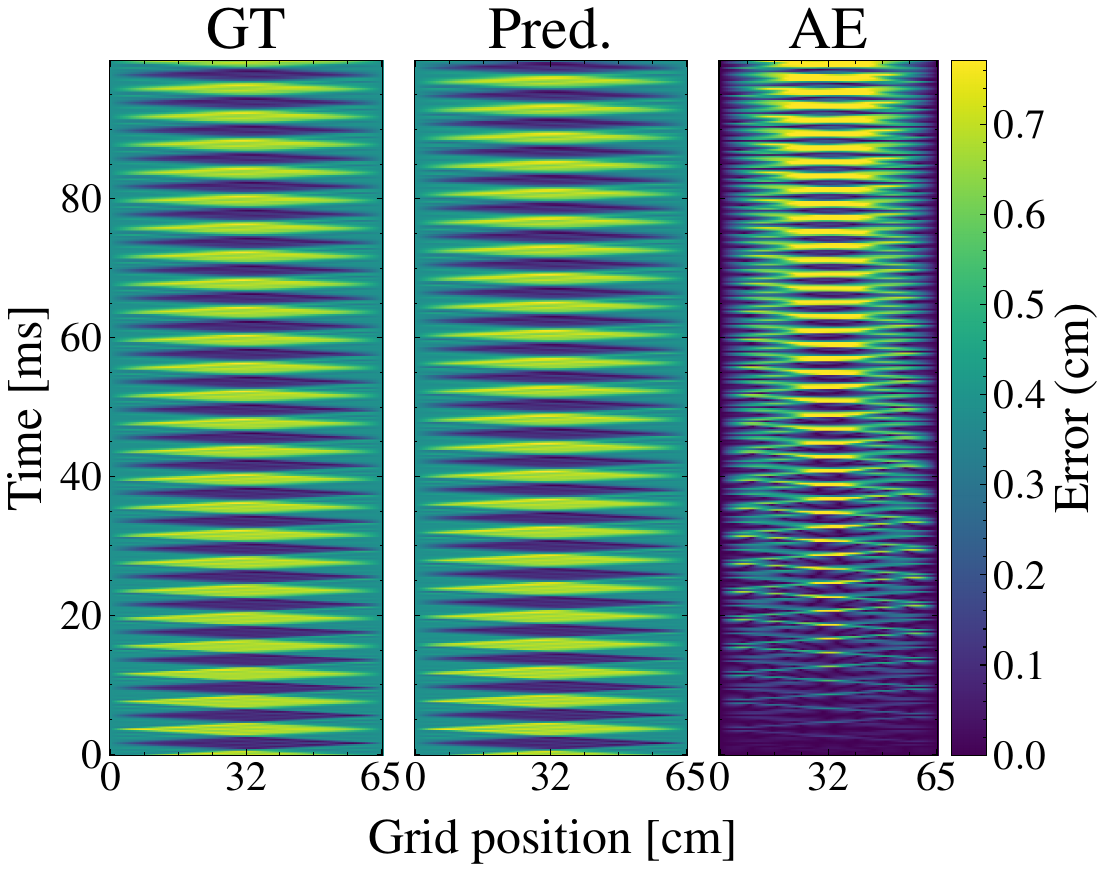} \\
            \im{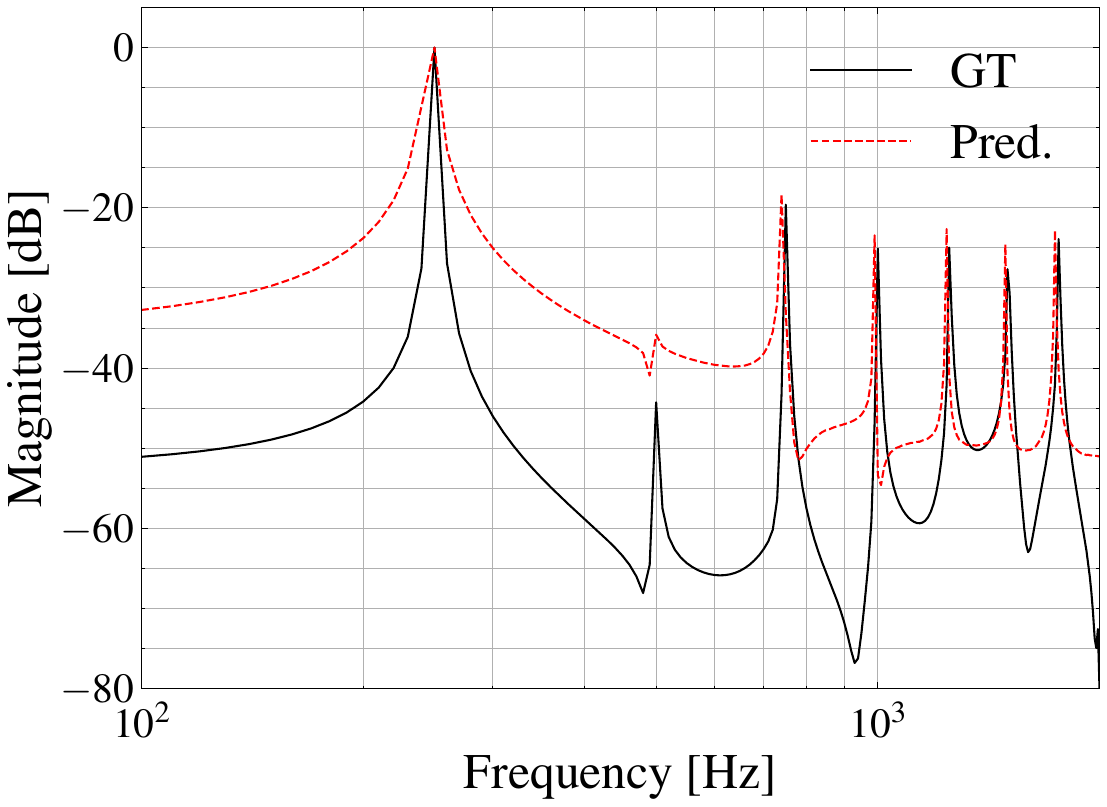} \\
            \small (a) DMD \\
        \end{tabular} &
        \begin{tabular}{c}
            \im{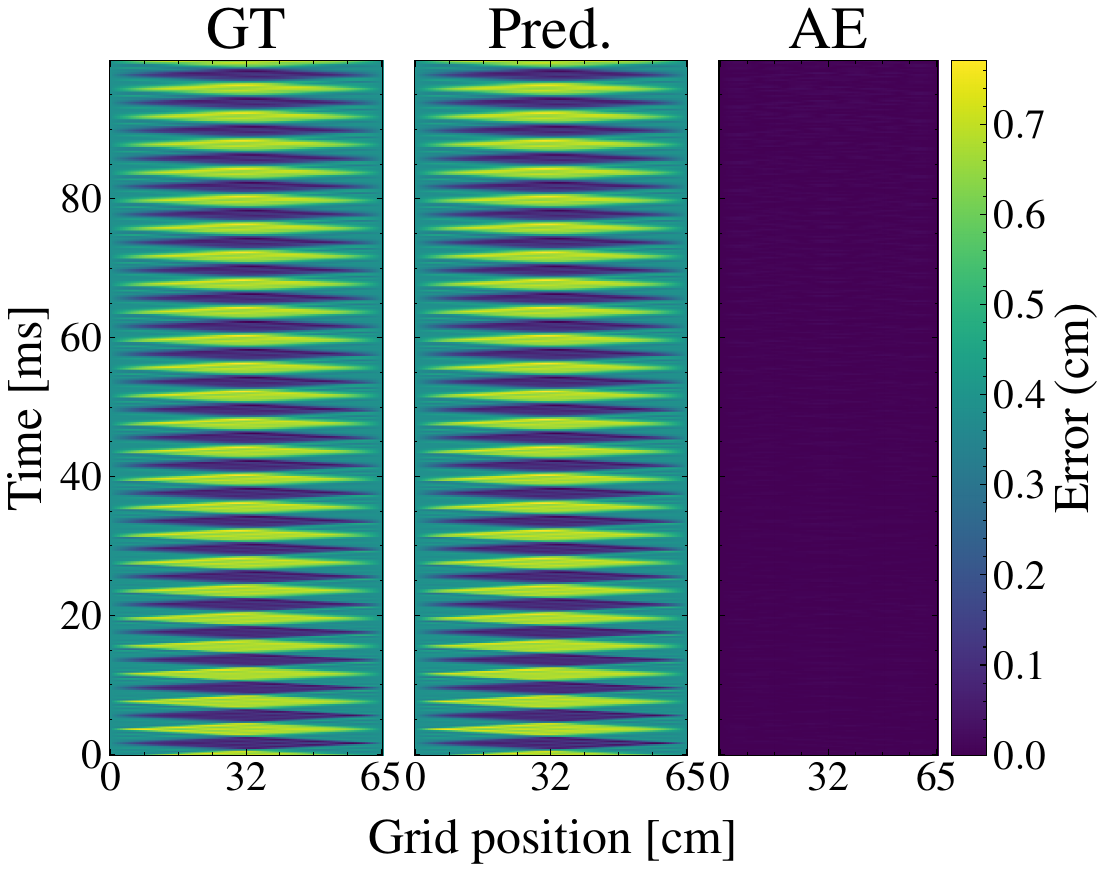} \\
            \im{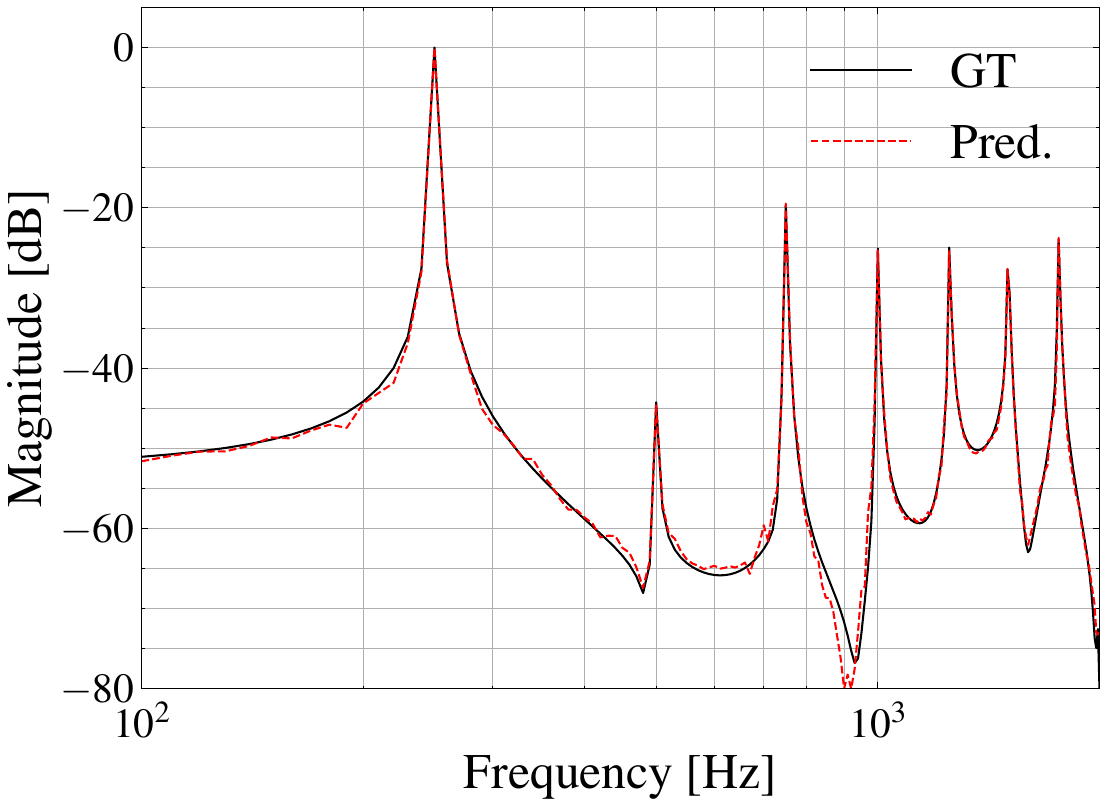} \\
            \small (b) Koopman\textsubscript{var} \\
        \end{tabular} &
        \begin{tabular}{c}
            \im{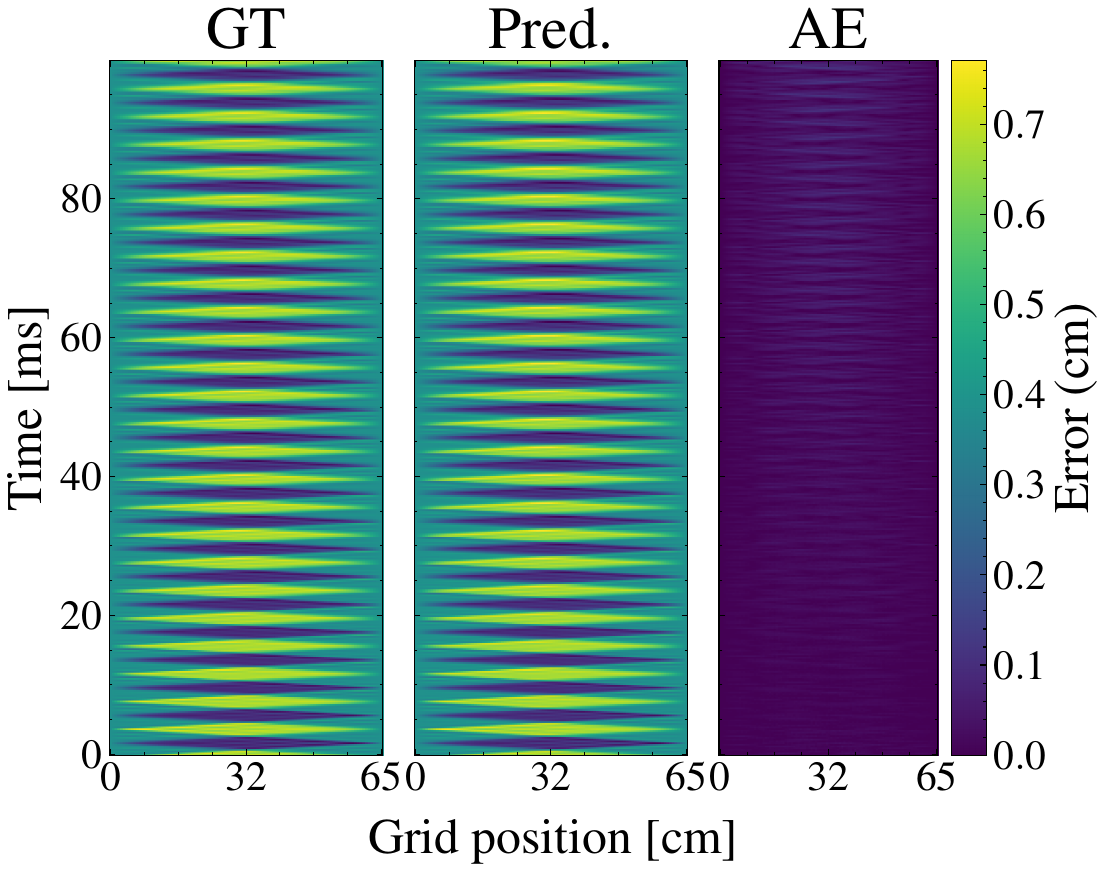} \\
            \im{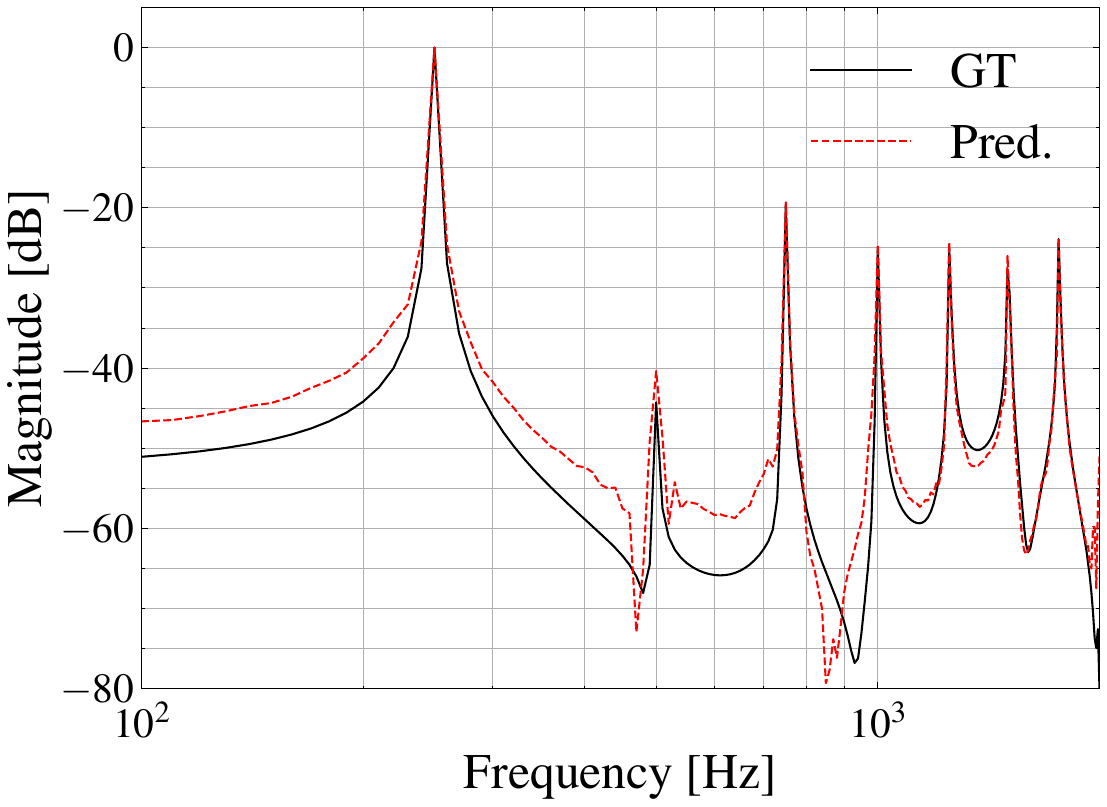} \\
            \small (c) Koopman \\
        \end{tabular} \\
    \end{tabular}

    \vspace{0.5em} % Adjust space between rows as needed

    \begin{tabular}{cccc}
        \begin{tabular}{c}
            \im{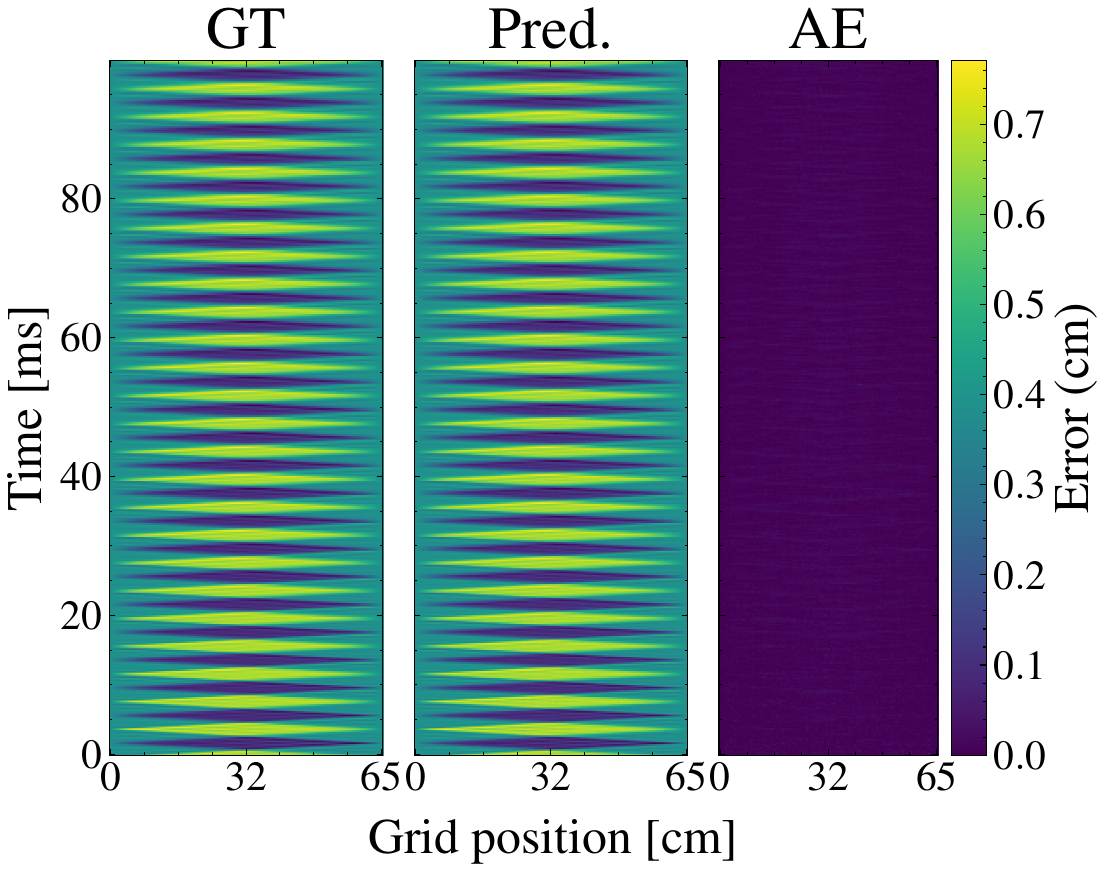} \\
            \im{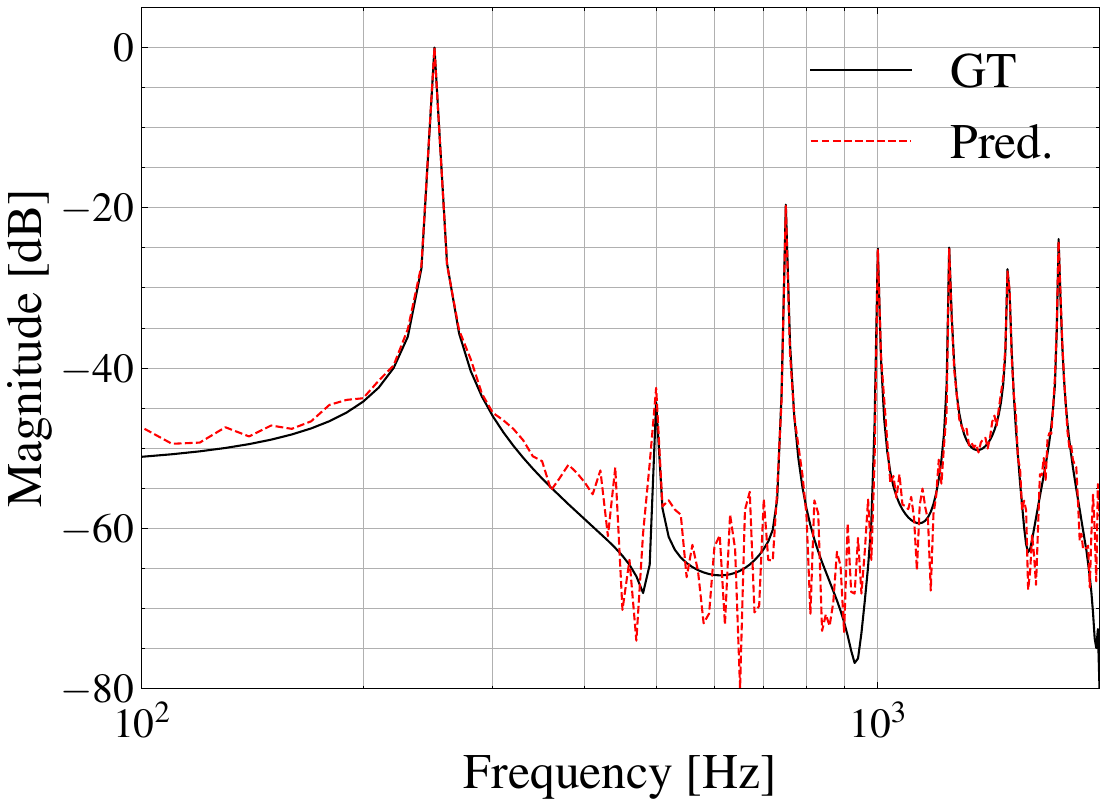} \\
            \small (d) LRU \\
        \end{tabular} &
        \begin{tabular}{c}
            \im{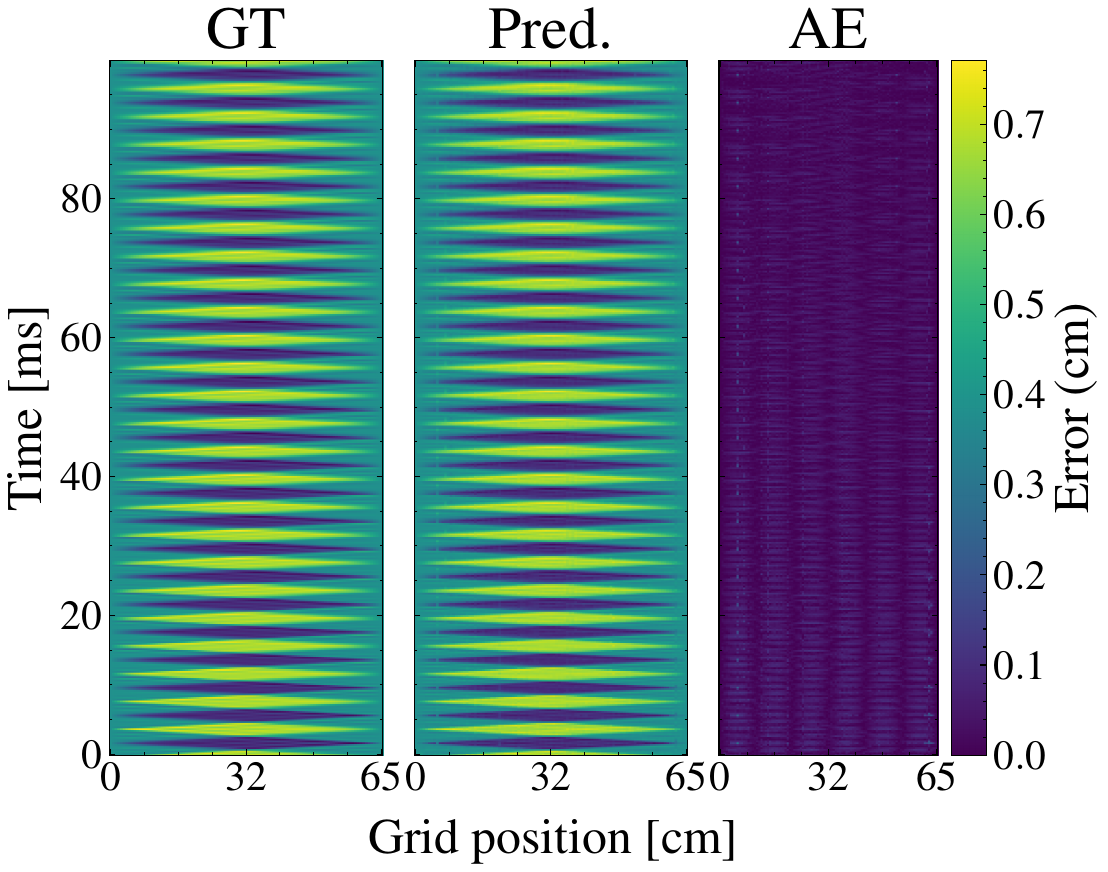} \\
            \im{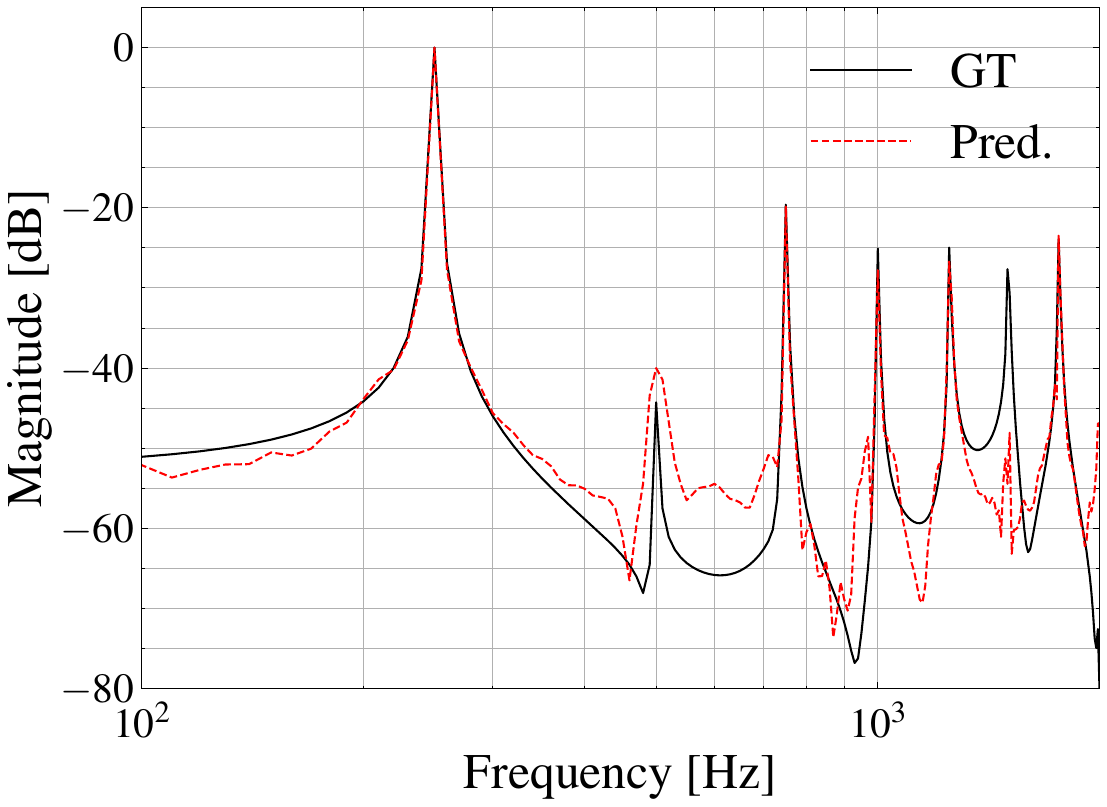} \\
            \small (e) S5 \\
        \end{tabular} &
        \begin{tabular}{c}
            \im{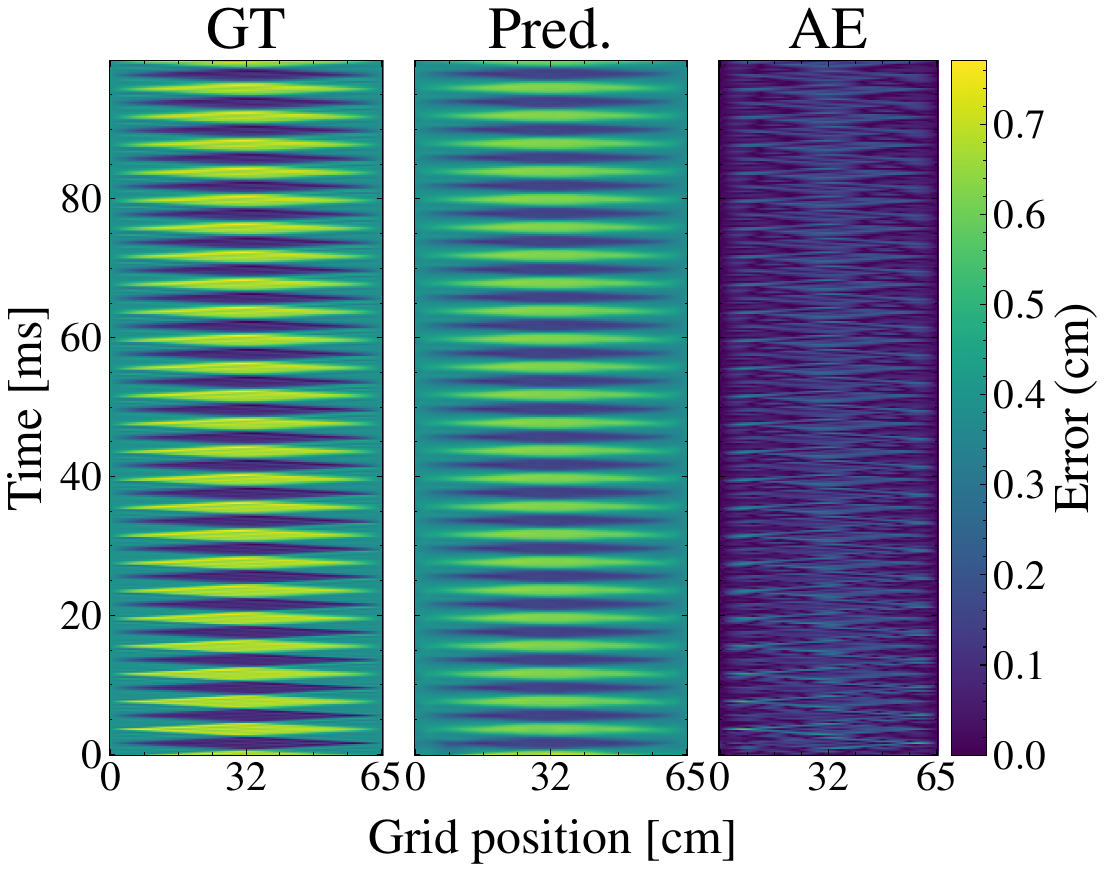} \\
            \im{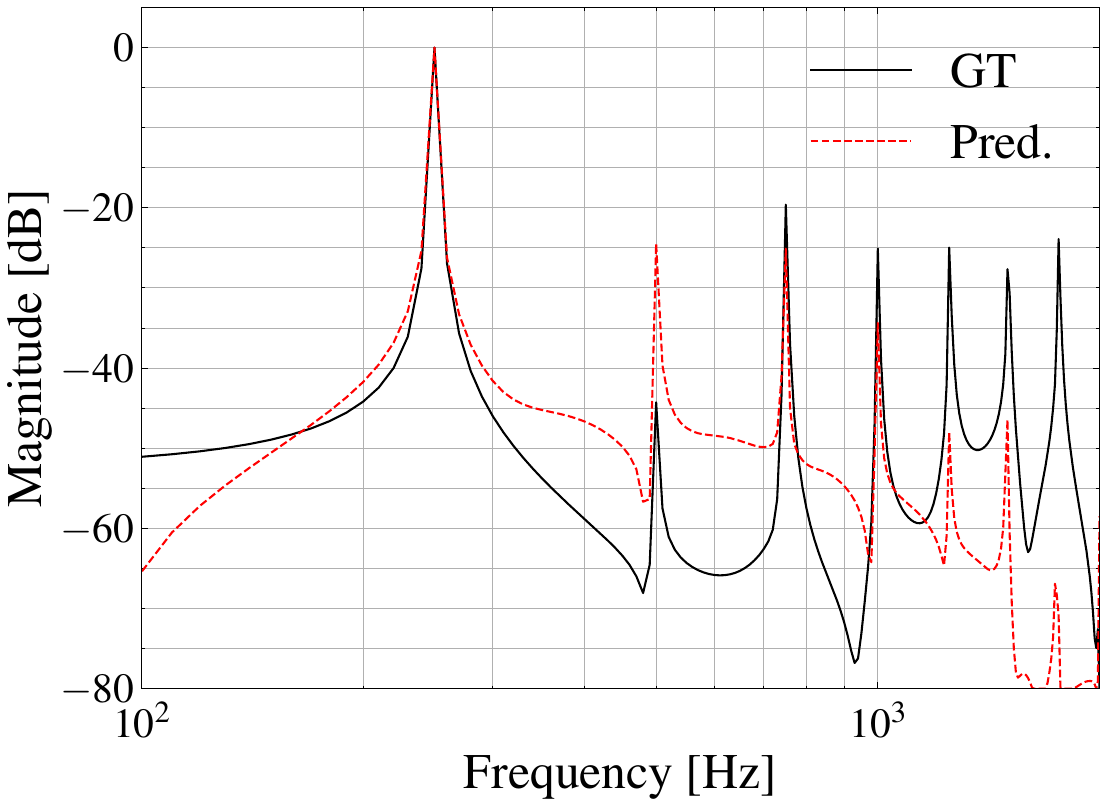} \\
            \small (f) FGRU \\
        \end{tabular} &
        \begin{tabular}{c}
            \im{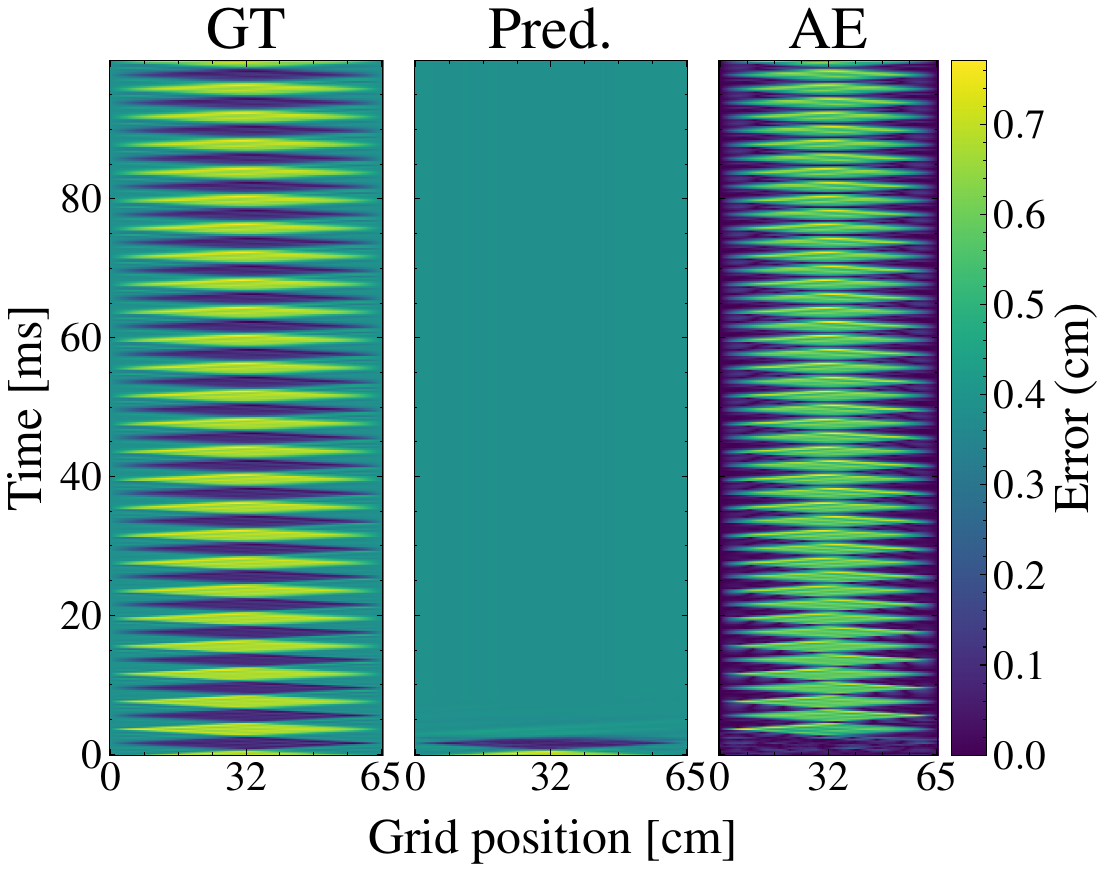} \\
            \im{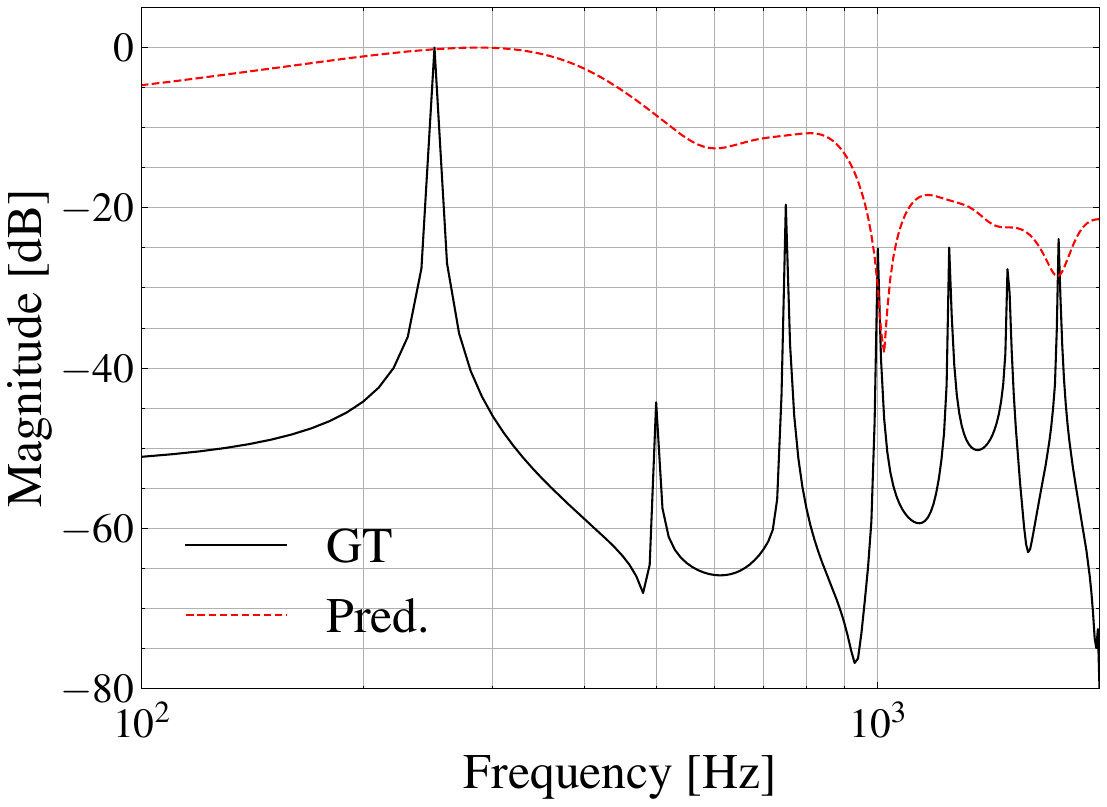} \\
            \small (g) FRNN \\
        \end{tabular}
    \end{tabular}
    \vspace{-4pt}
    \caption{Non-linear dynamics for an unseen uniform noise-like initial condition in the range 0 to 1\si{\cm}. The top row for each model shows the displacement evolution along the string for 400 time steps (100\si{\ms}). The bottom row for each model displays the spectrum of the same section at a single point ($\approx24\si{\cm}$). All models were trained on 400 time steps from the same dataset.}
    \label{fig:comparison_error_paper}
\end{figure*}

The results of our experiments are summarised in Tables~\ref{tab:results} for nonlinear datasets and~\ref{tab:results_linear} for linear datasets, showing the relative mean absolute error (MAE) and the relative mean square error (MSE). Additional results and visualisations can be found in the accompanying website~\footnote{\url{https://rodrigodzf.github.io/physmodjax/results.html}}. We use relative metrics, normalised with respect to the norm (squared in the case of MSE) of the string displacement trajectory.

In addition, we present metrics for the DMD results as a baseline. It is important to note that DMD struggles to capture the dynamics of systems such as ours, characterised by standing waves, due to the problem of linear consistency across consecutive time steps~\cite{tu_dynamic_2014}. To address this limitation, we augment the input data with a Hankel matrix constructed from two lags of the input~\cite{arbabi_ergodic_2017}. Consequently, this comparison might seem skewed, since the other methods effectively work with fewer data, making the evaluation somewhat unequal. Additionally, the DMD only fits a single sequence. To ensure a reasonable fit, we use a fixed amplitude of 0.0055 which is the mean of the variation in the amplitude of the training data and set the singular value rank to $r = 50$. We observe that using $r > 50$ singular value degrades the accuracy of the DMD prediction.

Our findings indicate that all models, except for the FRNN model, demonstrate the ability to capture the system's dynamics, though their accuracy varies. In particular, the Koopman\textsubscript{var} model recorded the lowest relative MAE in non-linear scenarios. The predicted displacement evolution of the model closely matches the ground truth data, as shown in Figures~\ref{fig:comparison_error_paper} and~\ref{fig:extrapolation_paper}. This accuracy is also reflected in the frequency content of the sequence. In the linear case, DMD performs better than the neural models. This is expected, as the optimal solution to the linear case can be achieved through the linear fitting of the DMD.

The FNO models perform worse than the other models (trained with 400 time steps), with the FRNN being significantly unstable during training and inference. Figure~\ref{fig:comparison_error_paper} shows the inference of the FNO models at their last stable checkpoint with the lowest validation MAE (approximately 500 epochs for the FRNN model).

\vspace{-4pt}
\subsubsection{Extrapolation in Time}
\vspace{0cm}

\begin{figure}[ht]
\begin{tabularx}{\textwidth}{cc}
\multicolumn{2}{c}{\includegraphics[width=\linewidth,clip]{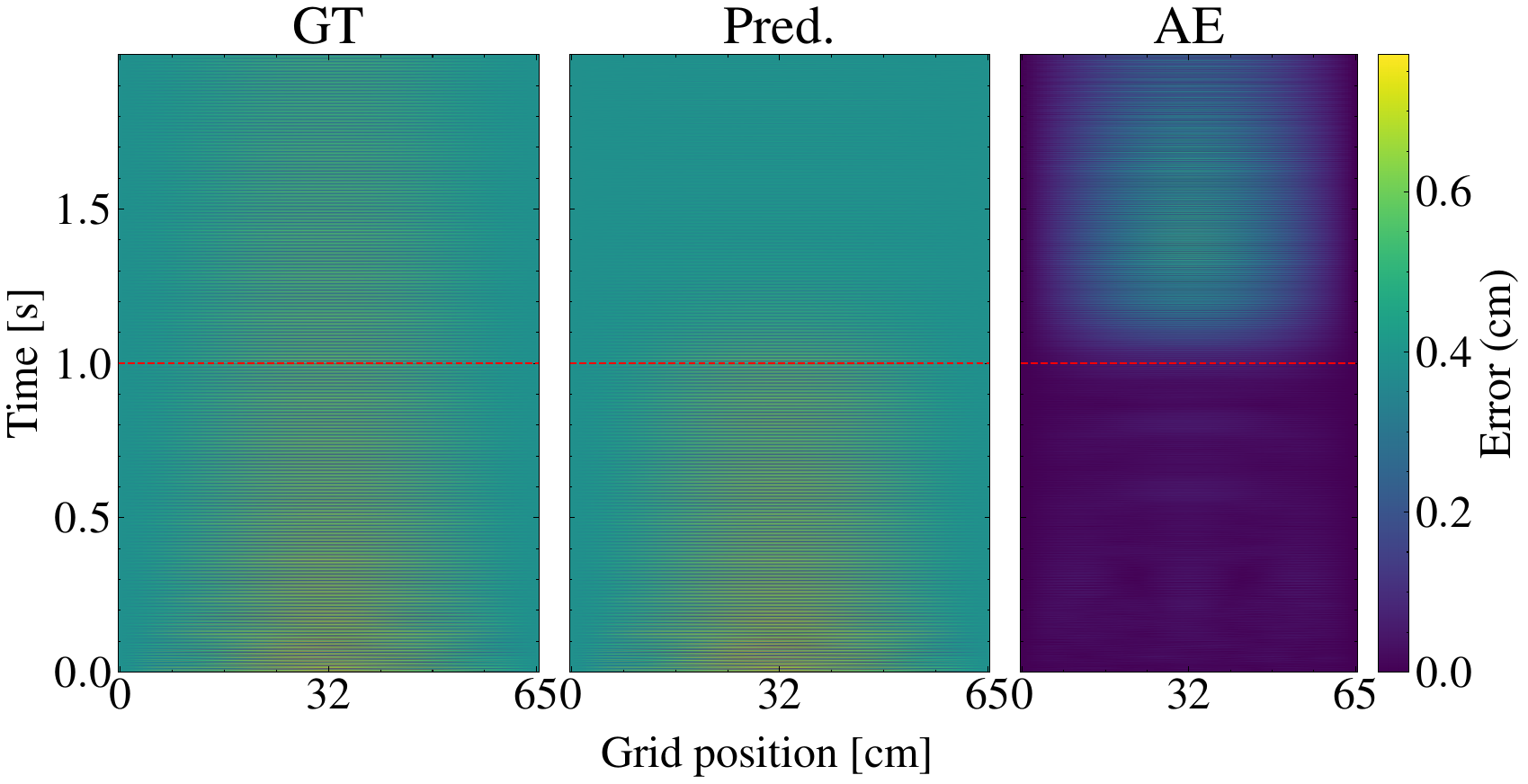}} \\
\includegraphics[width=0.45\linewidth,clip]{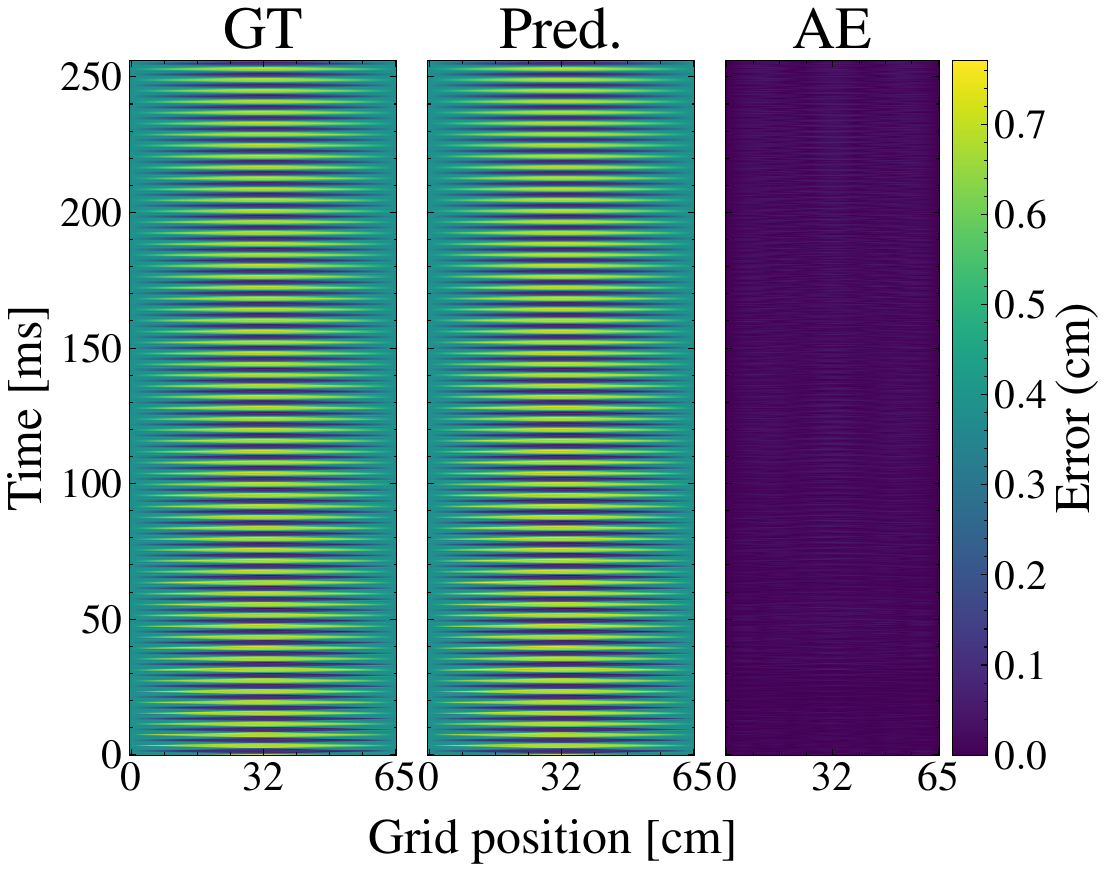}   &  \includegraphics[width=0.45\linewidth,clip]{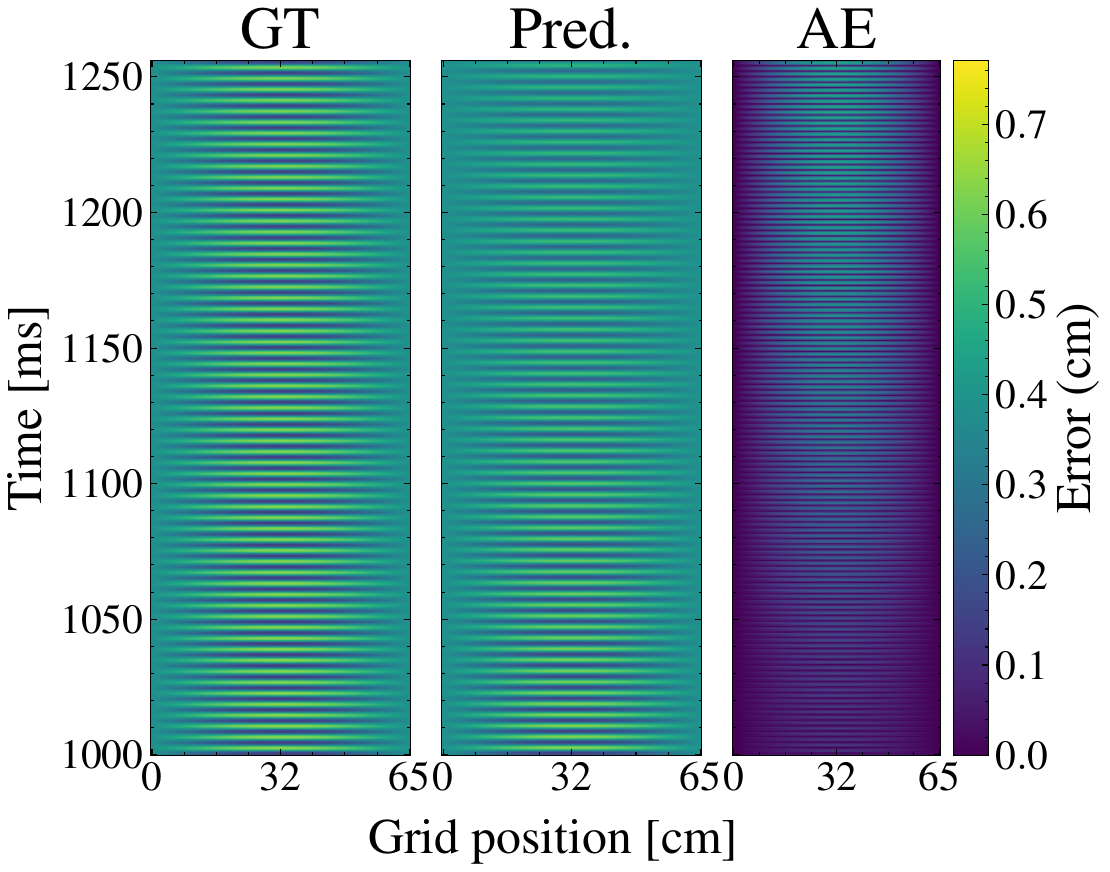}         \\
\includegraphics[width=0.45\linewidth,clip]{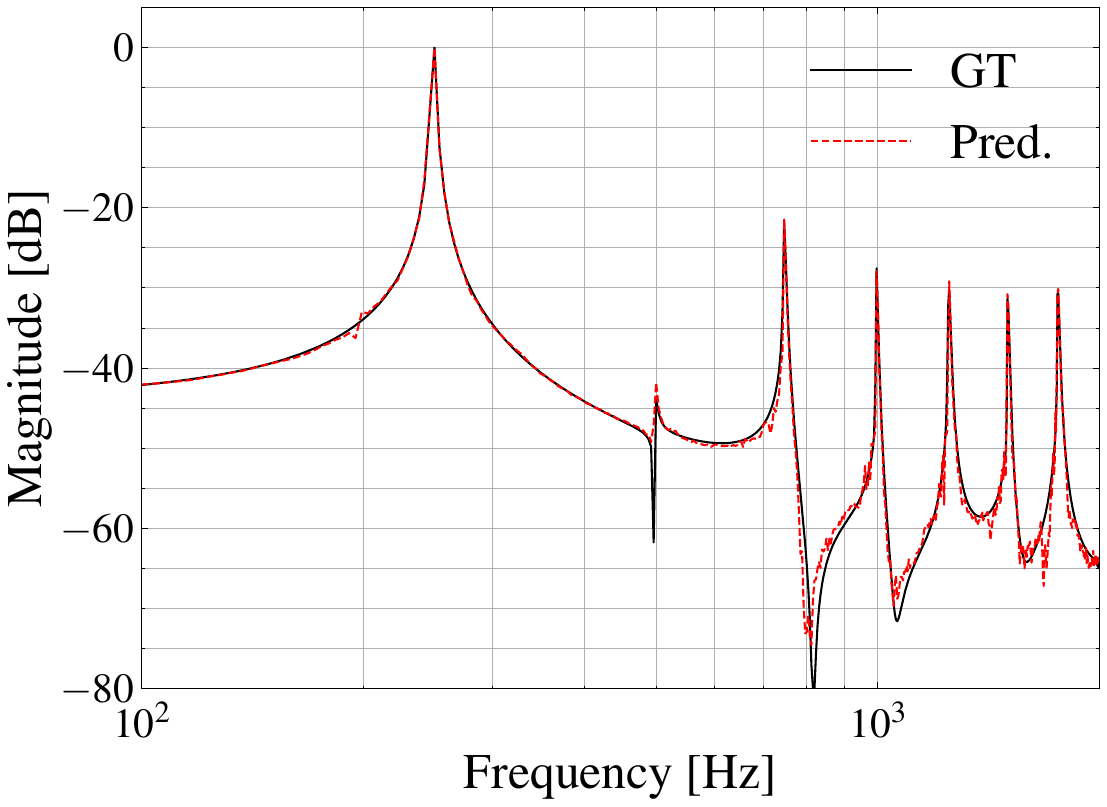}            & \includegraphics[width=0.45\linewidth,clip]{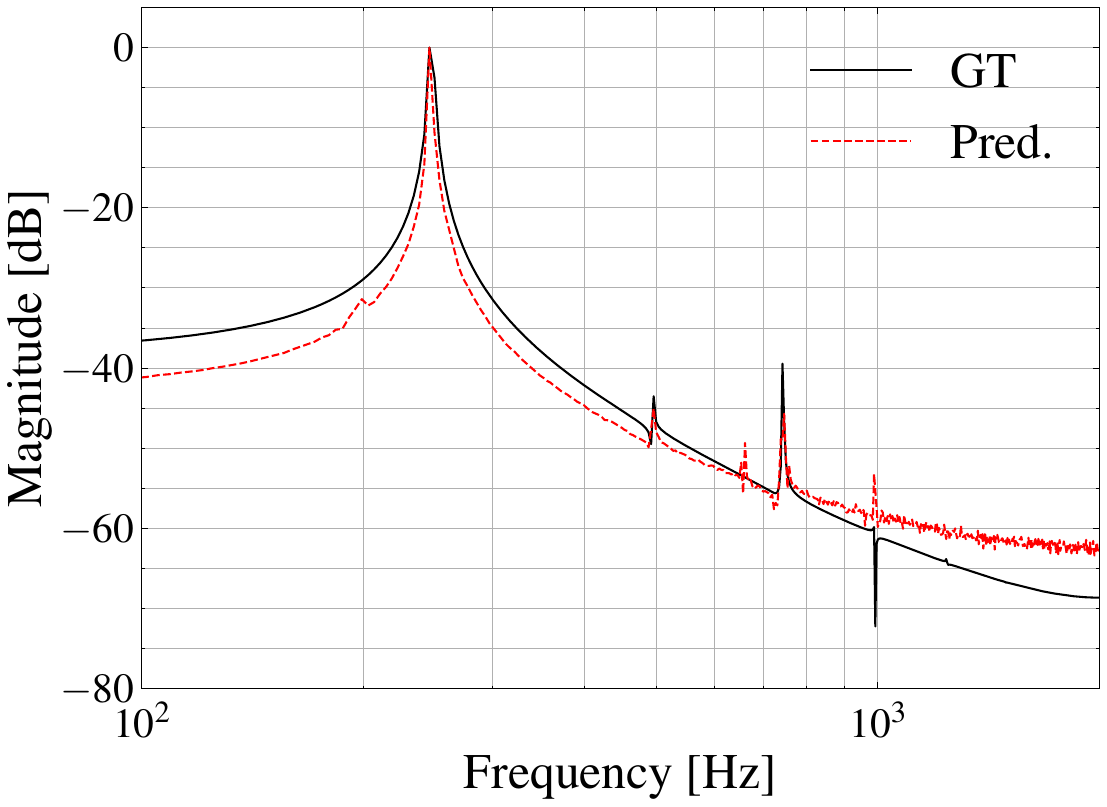}  
\end{tabularx}
\vspace{-8pt}
\caption{Evolution of the predicted displacement and extrapolation beyond the training horizon for a uniform noise-like initial condition (0 to 1\si{\cm}). The top row shows the Koopman\textsubscript{var} extrapolation with absolute error (in centimetres). The middle plots zoom in on the first 0.25 seconds (left) and 0.25 seconds after the training horizon (right). The bottom plots show the corresponding spectrum at a single position on the string ($\approx24$ cm) for the time spans covered in the middle plots.}
\label{fig:extrapolation_paper}
\end{figure}

\begin{figure}[ht]
\begin{subfigure}[b]{0.95\linewidth}
   \includegraphics[width=1\linewidth]{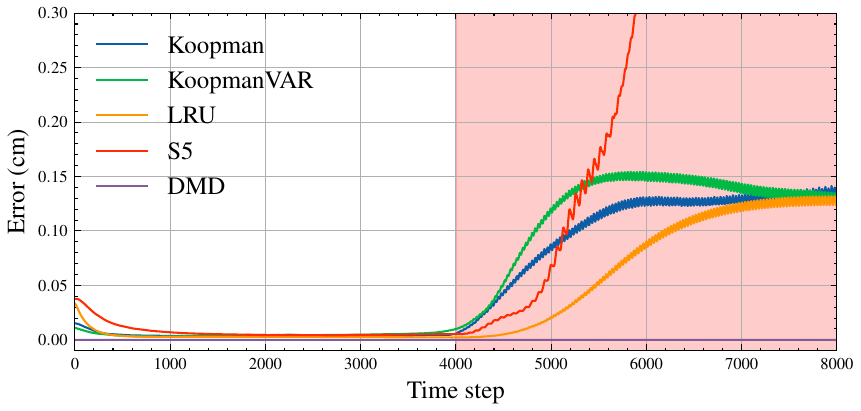}
   \caption{}
\end{subfigure}
% \hfill
\begin{subfigure}[b]{0.95\linewidth}
   \includegraphics[width=1\linewidth]{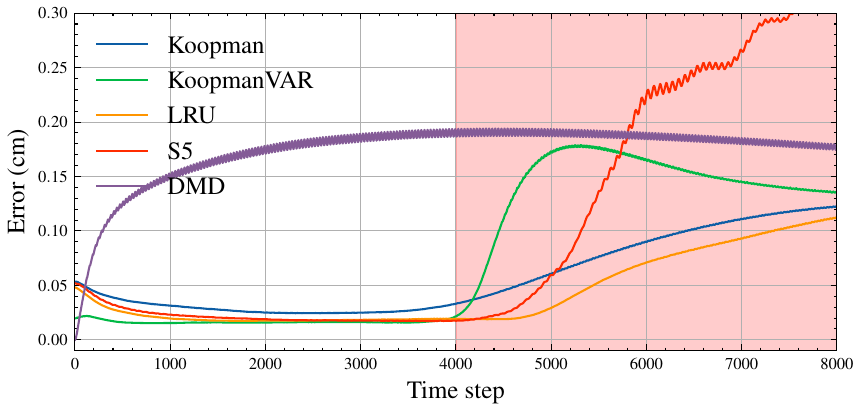}
   \caption{}
\end{subfigure}

\caption{Mean absolute error in centimetres per timestep across 100 unseen test trajectories for each architecture for 4000 time steps. Results are shown for a random noise-like initial condition (0 to 1\si{\cm}) (a) for a linear string and (b) for a tension-modulated string at a single position on the string ($\approx24$ cm). The regions shaded in red represent the predictions after the training time step horizon. The results for the FGRU and FRNN models are not included as these could not be trained with more than 400 steps. The error curves are smoothed for easier visualisation.}
\label{fig:mae_timestep}
\end{figure}

Although the models can generalise across different initial conditions, their capability in extrapolating beyond the predefined training horizon of 4000 time steps is limited. As depicted in Figure~\ref{fig:mae_timestep}, the models' predictions start to significantly deviate from the ground truth beyond the final training step. As expected, an exception exists with the DMD's performance in linear scenarios, where it provides the optimal linear solution. The predictions made by the Koopman\textsubscript{var} model begin to diverge from the actual data more rapidly than those made by the SSM models, which demonstrate slightly better accuracy for a marginally longer duration. In the case of the S5 model, we did not clip the eigenvalues during training, resulting in more unstable long-term inference. The divergence in the Koopman\textsubscript{var} model is mostly present in the higher frequencies and generally manifests as a shorter decay, as shown in Figure~\ref{fig:extrapolation_paper}.

This limitation in temporal extrapolation likely stems from the inherent non-linearities within the models and the exponential parameterisation used for their stability. These features are beneficial for learning dynamics within the observed timeframe and for enforcing stability, but they become problematic for predictions that extend beyond the trained interval. Unlike the FGRU or FRNN models, the Koopman-based models and SSMs remain mostly stable in the extrapolated region.

It is important to note that the models were trained exclusively on displacement data; velocity information was not provided as input in our experiments. Representing the state of the system solely through displacement or its embedding might be insufficient for an unequivocal state description.

%% file: 05_conclusion.tex
\vspace{-4pt}
\section{Conclusions}
\label{sec:conclusions}
% \vspace{-4pt}

We have explored SSM and Koopman-based methods for modelling of the dynamics of linear and tension-modulated strings, providing a comprehensive comparison with similar methods. Our experimental results highlight the viability of Koopman-based methods in effectively modelling these systems over extended time horizons, though challenges remain.

A significant finding from our research is the better or at least comparable performance of the state-varying Koopman method compared to the SSM and FNO models in predicting the dynamics of the string. This finding underscores the potential efficiency of Koopman-based models, which, despite having fewer parameters~\ref{tab:parameters}, may be better suited to such dynamic systems.

Looking ahead, our findings open up several avenues for further investigation. We believe that it might be possible to mitigate the issue with extrapolation in time either by augmenting the input (e.g., by providing two concatenated time steps at each step, similar to the Hankel DMD approach~\cite{arbabi_ergodic_2017}) or by incorporating initial velocity information. The latter strategy, in particular, could equip the models with the necessary information to initiate dynamic evolution from a later state, akin to starting from a new initial condition. The absence of velocity data means missing essential phase information needed for long-term state predictions from subsequent starting points.

Another challenge to address is adapting our models to different physical parameters. Currently, the model learns the trajectory for a single set of physical parameters. However, the Koopman and SSM models could be adapted with an additional layer that encodes not only the initial condition but also the corresponding physical parameters. Additionally, we intend to explore the interpretability of the learned weights in the Koopman model. It may also be possible to directly manipulate these weights for more creative uses.
Expanding the application of these modelling techniques to two and three dimensional domains represents another possible extension of our method. Such efforts could test the adaptability of our methods to more complex problems in multidimensional systems.

In summary, our exploration of SSMs and Koopman-based modelling techniques reveals a promising landscape for future research in simulating the dynamics of acoustical systems. By addressing the identified limitations and exploring proposed future directions, we can continue to expand the boundaries of what is possible in physical modelling with neural networks.

%% file: 06_acknowlegments.tex
\vspace{-4pt}
\section{Acknowledgments}\label{sec:acknowledgement}
\vspace{-4pt}
% Funded by UKRI and EPSRC as part of the ``UKRI CDT in Artificial Intelligence and Music'', under grant EP/S022694/1.
This research utilised Queen Mary's Apocrita HPC facility, supported by QMUL Research-IT. http://doi.org/10.5281/zenodo.438045